\documentclass[preprints,article,accept,moreauthors,pdftex]{Definitions/mdpi}
\firstpage{1} 
\pubvolume{1}
\issuenum{1}
\articlenumber{0}
\pubyear{2021}
\copyrightyear{2020}
\datereceived{} 
\dateaccepted{} 
\datepublished{} 
\hreflink{https://doi.org/} 
\pdfoutput=1

\usepackage{longtable}
\usepackage{lscape}
\usepackage{adjustbox}
\usepackage{rotating}
\usepackage{newtxtext,newtxmath}
\usepackage{graphicx}	
\usepackage{amsmath}	
\usepackage{amssymb}	

\Title{Discovery of 178 Giant Radio Galaxies in 1059 deg$^2$ 
      of the Rapid ASKAP Continuum Survey at 888 MHz}

\TitleCitation{Discovery of 178 Giant Radio Galaxies in 1059 deg$^2$ 
      of the Rapid ASKAP Continuum Survey at 888 MHz}

\Author{Heinz Andernach $^{1}$\orcidA{}*, Eric F. Jim\'enez-Andrade $^{2}$\orcidB{} and Anthony G. Willis $^{3}$\orcidC{}}

\AuthorNames{Heinz Andernach, Eric F. Jim\'enez Andrade and Anthony G. Willis}

\AuthorCitation{Andernach, H.; Jim\'enez-Andrade, E.F.; Willis, A.G.}

\address{%
$^{1}$ \quad Departmento de Astronom\'{i}a, DCNE, Universidad de Guanajuato, 
Callej\'on de Jalisco s/n, Guanajuato, C.P. 36023, GTO, Mexico; heinz@ugto.mx\\
$^{2}$ \quad National Radio Astronomy Observatory, 520 Edgemont Road, Charlottesville, VA 22903, USA; ejimenez@nrao.edu \\
$^{3}$ \quad 310 Yorkton Avenue, Penticton, British Columbia, Canada V2A 6Z8; tony.willis.research@gmail.com }
\corres{Correspondence: heinz@ugto.mx}

\abstract{We report the results of a visual inspection of images of the 
Rapid ASKAP Continuum Survey (RACS) in search of extended radio 
galaxies (ERG) that reach or exceed linear sizes on the order of one Megaparsec.
We searched a contiguous area of 1059\,deg$^2$ from
RA$_{\rm J}$=20$^h$20$^m$ to 06$^h$20$^m$, and $-50^{\circ}<\rm{Dec}_J<-40^{\circ}$, 
which is covered by deep multi-band optical images of the 
Dark Energy Survey (DES), and in which previously only three ERGs 
larger than 1\,Mpc had been reported. For over 1800 radio galaxy candidates 
inspected, our search in optical and infrared 
images resulted in hosts for 1440 ERG, for which spectroscopic and 
photometric redshifts from various references were used to convert their
largest angular size (LAS) to projected linear size (LLS). This
resulted in 178 newly discovered giant radio sources (GRS) with 
LLS$>$1\,Mpc, of which 18 exceed 2\,Mpc and the largest one is 3.4\,Mpc.
Their redshifts range from 0.02 to $\sim$2.0, but only 10 of the 178 
new GRS have spectroscopic redshifts. For the 146 host galaxies the 
median $r$-band magnitude and redshift are 20.9 and 0.64, while for 
the 32 quasars or candidates these are 19.7 and 0.75. Merging the six 
most recent large compilations of GRS results in 458 GRS larger than
1\,Mpc, so we were able to increase this number by $\sim39\%$ to
now 636.}


\begin{document}

\section{Introduction}  \label{intro}

Giant radio galaxies (GRG) were first discovered in 1974 \cite{willis74}
and originally defined as those for which their projected linear size
(LLS) exceeds 1\,Mpc, at a time when the Hubble constant was assumed to be
$H_0 = 50$\,km\,s$^{-1}$\,Mpc$^{-1}$. Today $H_0$ is accepted to be 
near 70\,km\,s$^{-1}$\,Mpc$^{-1}$ so that objects of the same angular 
size now have a linear size smaller by a factor
of $\sim$1.4. Thus the first GRG found (3C\,236), originally assigned
an LLS of 5.7\,Mpc, is now considered to extend over 4.1\,Mpc.
This change in the adopted value of $H_0$ has led more recent authors
to adopt an LLS of 0.7\,Mpc as a lower limit for considering
a radio galaxy a ``giant'' one. However, there is no physical reason for 
adopting either of these thresholds, and in the present 
paper, we focus on only those larger than 1\,Mpc.

In 1983 \cite{hintzen83} the first two giant radio quasars (GRQ) 
were found, [HB89]~1146--037 of LLS=1.06\,Mpc, at a redshift of z=0.341, and
[HB89]~1429+160 with 1.37\,Mpc at z=1.016, and these authors already 
raised the issue that inverse Compton losses suffered by the
synchrotron-emitting electrons by scattering the cosmic
microwave background (CMB) photons at $z\gtrsim1$, and the
denser Universe at that cosmic epoch, should not let these GRQ
expand to the same extent as in the local Universe. Since then
many other GRQs have been found, and currently they constitute 
about 18\% of all known giant radio sources (GRS), see also 
Sect.~\ref{sect:compar}. Unless otherwise noted, here
we use the terms GRS, GRG, and GRQ for sources with LLS$>$1\,Mpc.

Despite an intense search for further and larger examples of these rare 
objects, only two GRGs larger than 3C\,236 have been found so far, namely
J1420--0545 of 4.7\,Mpc \cite{machalski08} and J0931+3204 of 4.3\,Mpc 
\cite{coziol17}, the latter being the largest GRQ currently known. 
A third GRG of supposedly 4.45\,Mpc, listed as J1234+5318 in the
latest GRS compilation by \cite{kuzmicz18}, was originally published by
\cite{bandfield15}, but was later shown to have a lower-redshift host
on the basis of a deeper image from the Low Frequency
Array (LOFAR) \cite{osullivan19}, which led to a smaller LLS of 3.25\,Mpc. 
This latter
object, with a largest angular size (LAS) of $\sim11'$ is an interesting
example of the difficulty to correctly identify the host of a radio
galaxy with widely spaced lobes.

As shown by \cite{willis74, bridle76}, while GRGs are not the most radio 
luminous extragalactic radio sources, they tend to have the
lowest minimum energy densities (down to $\sim10^{-14}$\,J\,m$^{-3}$)
in relativistic particles and magnetic field, and, due to their huge volume,
they have the largest total energy content in particles and magnetic field,
up to a few 10$^{54}$\,J, and as such, as expressed by \cite{strom74}, 
``put the greatest strain on radio source models''.

Until recently the search for GRS was mainly based on two large-scale
radio surveys, the NRAO VLA Sky Survey (NVSS, \cite{condon98}) with
an angular resolution of 45$''$ and sensitive to the 
diffuse emission from radio galaxy lobes, and the Faint Images of the 
Radio Sky at Twenty centimeters (FIRST, \cite{becker95}) which
due to its higher angular resolution of $5.4''$ best reveals
the radio nuclei and thus the host position of extended sources.
Both  surveys were carried out at 1.4\,GHz with the Very Large
Array (VLA) in New Mexico, USA, which has led to the fact that the 
majority of known GRS were found in the northern celestial
hemisphere, which is also largely covered by deep imaging and 
spectroscopic surveys like the Sloan Digital Sky Survey (SDSS, \cite{sdssdr16}). 
Thus, the latest literature compilation \cite{kuzmicz18} lists 213 GRS, 
of which 20 (9.4\%) lie south of Dec$=-40^{\circ}$, an area of 17.9\%
of the sky. Only recently, with the advent of sensitive low-frequency
surveys like the LOFAR Two-metre Sky Survey (LoTSS, \cite{shimwell19})
and those performed with the Australian Square Kilometre Array Pathfinder
(ASKAP, see \cite{hotan21}) more GRS are being found in both 
hemispheres \cite{dabhade20a, brueggen21, norris21b}.

In order to increase the number of known GRS especially in the
far southern hemisphere (Dec$<-40^{\circ}$)
we explore here the possible content of GRS in the recent
Rapid ASKAP Continuum Survey (RACS, \cite{mcconnell20}) which has
imaged 83\% of the entire sky (Dec$<+41^{\circ}$) at 888\,MHz with 
an angular resolution of 15$''$ and an average 
noise level of 0.25~mJy\,beam$^{-1}$. Both  characteristics 
are three times better than that of the previously best southern radio
survey, the Sydney University Molonglo Sky Survey (SUMSS, \cite{bock99, mauch03}).
This makes RACS the first radio survey to cover the far southern sky 
that allows to identify the hosts of the majority of extended radio sources 
in modern optical/IR surveys like SkyMapper (SMSS, \cite{wolf18}),
the Dark Energy Survey (DES, \cite{abbott18}) and those of the Wide-field 
Infrared Survey Explorer (WISE, e.g.\ \cite{cutri13, lang14}).

As already found by \cite{komberg09}, the often cited reasons for the 
extreme sizes acquired by GRS, namely (i) a preferred orientation
in the plane of the sky, (ii) a location in lower-density
environments, and (iii) more powerful jets feeding their lobes,
cannot explain the observations, and these authors argue that a more
likely reason is a much longer duration of their radio active phase,
although the latter is difficult to prove. Further 
motivations for a continued search for GRS are (a) the question 
of whether there is a physical limit near $\sim$5\,Mpc for GRS, and 
(b) to use them as probes of the large-scale structure of the Universe,
since \cite{malarecki13, malarecki15} have shown, albeit on 
a limited sample of relatively nearby GRS, that the major axis
connecting the two lobes of GRS have a tendency to lie perpendicular
to the orientation of the ambient galaxy overdensity or that
of galaxy filaments of which the host is a member. A list of
extended RGs of any size, but with a high surface density per
deg$^2$ would also be useful to test conjectures of alignments of
their radio position angles like those claimed by \cite{taylor16} and 
\cite{Contigiani17}. Eventually we were also interested in 
finding out how well a relatively low angular resolution
survey like RACS and the imminent all-sky survey ``Evolutionary 
Map of the Universe" (EMU, \cite{norris21b}) would be suited to
unambiguously identify the hosts of extended radio sources.

In this paper we adopt standard cosmological parameters of
$H_0=70$\,km\,s$^{-1}$\,Mpc$^{-1}$, $\Omega_{m}$=0.3, $\Omega_{\Lambda}$=0.7,
and a radio source spectral index $\alpha$ defined as
$S_{\nu}\propto{\nu}^{\alpha}$.

\section{Methodology}  \label{methods}

Given that  \cite{boyce00} had screened the then
existing SUMSS images for GRS in an area of 2100\,deg$^2$ defined
by Dec${\rm_J}<-50^{\circ}$ and Galactic latitude $|b|>12.5^{\circ}$) 
and had found 21 objects with LAS$>5'$, later published in
\cite{saripalli05}, we selected for the first systematic search
for GRS in RACS the declination range from $-40^{\circ}$ to $-50^{\circ}$.
Knowing that radio galaxies tend to be found up to high redshifts,
implying very faint hosts,
we further limit ourselves to the high Galactic latitude area from
RA=20$^h$20$^m$ to 06$^h$20$^m$ (a total of 1059\,deg$^2$) as it is
covered by DES, providing deep optical $g,r,i$-band images and photometric 
redshifts. The literature compilation by \cite{kuzmicz18} lists only three
GRS in this area.

Small parts of this area had been surveyed previously in radio,
and the corresponding images had been inspected by one of us (H.A.)
for the presence of GRS. Firstly, the
Australia Telescope ESO Slice Project (ATESP, \cite{prandoli00}),
covering RA${\rm_J}$=22$^h$32$^m$--22$^h$57$^m$ and 23$^h$31$^m$--01$^h$23$^m$ in
the range $-39.5^{\circ}<\rm{Dec_J}<-40.4^{\circ}$, overlaps for 
$\sim$17\,deg$^2$ with our search area.
Secondly, a circular area of $\sim$4.5\,deg$^2$ had been imaged with 
the Australia Telescope Compact Array (ATCA) at 1.4\,GHz as the 
Phoenix Deep Field (PDS) \cite{hopkins98,hopkins03} centered on 
RA${\rm_J}$,Dec${\rm_J}$=01$^h$13$^m$36$^s$,$-45.7^{\circ}$. 
Thirdly, an area of $\sim40$\,deg$^2$ 
(RA${\rm_J}$=20$^h$05$^m$--22$^h$26$^m$,$-48.1^{\circ}<\rm{Dec_J}<-50^{\circ}$)
overlaps with our search region and has been imaged with 
ASKAP as part of the EMU Pilot Survey \cite{norris21b}
at the same angular resolution as RACS, but to a 1-$\sigma$ noise 
level 10 times lower ($\sim26{\mu}$Jy\,beam$^{-1}$). We include
in our results only those GRS that can be recognized on RACS
images.

\subsection{Images used}

For our visual inspection we used the RACS images for Stokes~I data release~1, available in 
full resolution from
\url{https://www.atnf.csiro.au/research/RACS/RACS\_I1}.
The beam size and shape of the latter images vary with sky position
\cite{hale21}, such that the beam major axes range from 15$''$ to $25''$. 
The RACS Stokes~I data release~1 images are also provided as ``CRACS", 
convolved to a common resolution of 25$''$ at the URL \url{https://www.atnf.csiro.au/research/RACS/CRACS_I2},
which were used to prepare the source catalogue in \cite{hale21}. 
Both versions are in J2000 coordinates and pixel
intensities are calibrated in Jy\,beam$^{-1}$. For our search 
for GRS we preferred the full-resolution images to better constrain the 
most likely host position, but occasionally consulted CRACS
to confirm the presence of diffuse emission.
The astrometric precision of RACS is better than 1$''$ \cite{mcconnell20,hale21}
and thus does not affect the reliability of optical identifications.
It is rather the angular resolution that causes an increased rms of the
source position which amounts to $\sim2.7''$ for a faint source of S/N$\sim5$
\cite{mcconnell20}.

During a summer internship in 2012 at Univ.\ of Guanajuato one of us 
(E.F.J.A.), together with R.F.\ Maldonado S\'anchez \cite{andernach12}
had inspected all of the SUMSS images, logging the 
positions of $\sim$5000 potentially extended radio sources. Since then, 
a fraction of these had been followed up by one of us (H.A.) 
to either optically identify their host, or discard them as 
unrelated sources. However, the low angular resolution of SUMSS (45$''$),
together with the limited sensitivity of the optical Digitized Sky
Survey \cite{lasker96} often made this task a guesswork, and only the 
advent of RACS in radio and DES in the optical promised a solution.
Thus, in the first round of our search for GRS, we inspected RACS images 
at a few hundred previously logged positions in our search area. 
We used {\sc Aladin} (\cite{bonnarel00}, \url{http://aladin.u-strasbg.fr/aladin.gml}) 
to display 
these images, adjusting the contrast such that the noise floor could always 
be recognized. Whenever the source structure in RACS appeared as two or
more unrelated sources, they were discarded, and otherwise the 
$g,r,i$-composites of optical images from DES were consulted to find the 
most likely host, and its position was recorded for later retrieval of
complementary data. Over 260 radio sources spotted in SUMSS were thus
identified with the help of RACS.

In the second round, we systematically screened the full area of
1059\,deg$^2$ in the full-resolution RACS, marking on it with symbols
those sources already identified in the first 
round, as well as apparently extended sources that had been discarded 
previously, so as to avoid repeating the identification process for these.

Given that our aim was to find, apart from GRS also those with an appreciable
size of LLS$\gtrsim$0.5\,Mpc, we tried to log all possible sources
with a LAS$\gtrsim1.0'$, since for our adopted cosmology this is 
the minimum angular size a standard ruler of 0.5\,Mpc would appear 
if it had a redshift in the range of $\sim$1.4--1.8 (see e.g.\ Fig.\ 30 
on p.\ 1326 of \cite{carroll14} and
our $LAS-z$ diagram in Section~\ref{sect:compar}).
We also included, though less complete for LAS$\lesssim1.5'$,
bent-tailed radio sources like wide-angle tailed
(WAT) and narrow-angle tailed (NAT) sources, since these may serve as
indicators of the presence of a cluster of galaxies. 
We proceeded from RA=20$^h$20$^m$ eastward by scanning portions
of 1\,h wide in RA (or $\sim100$\,deg$^2$ at the chosen declination),
displaying about 1\,deg$^2$ per screen at a time and logging between 1 and
1.5 candidates per deg$^2$, which took about 16\,h for the entire 1059\,deg$^2$.
This resulted in a list of approximate centre positions of
$\sim$1330 candidate extended radio sources.

\subsection{Scrutinizing the Candidates: Host Identification and Radio Morphology}

The positions of all the $\sim$1330 candidates
selected in the previous step were then displayed again
in {\sc Aladin} and the likely host was searched on DES\,DR1
$g,r,i-$band color composites \cite{abbott18}. DES~DR1 has a median 
delivered point-spread function of 1.12, 0.96,
and 0.88$''$ in the $g,r$, and $i$-bands, an astrometric precision of
0.151$''$, and limiting magnitudes of 24.28, 23.95, and 23.34   
mag in $g,r$, and $i$-bands for objects with S/N=10 and a 2$''$-diameter aperture.
We also used mid-infrared (MIR) images from unWISE \cite{lang14}, colored 
according to the magnitude difference between the two lowest-wavelength WISE bands, $W1-W2$.  For sources with an obvious central radio core, finding the
host object was easy, both for the edge-darkened
Fanaroff-Riley \cite{fr74} type~I (or FR\,I) sources with a radio
brightness peak close to the optical host, and for the edge-brightened
FR\,II sources with a radio core. The problem
arises for sources with widely separated radio components, and without
an obvious central component in between them. 
If the central component had no optical or MIR counterpart, the pair of
outer components was discarded as a genuine RG. However, if one of the
outer components had a convincing host by itself and its radio extent
was larger than $\sim1'$, it was recorded as a separate extended RG.
If the central component had a clear optical or MIR host, the candidate
was discarded if the outer components had an obvious counterpart,
unless their radio structure showed clear indications of being connected
(e.g.\ radio bridges or trails pointing at each other) and the 
optical/MIR objects in the lobes did not show evidence for being 
active galactic nuclei (AGN). When felt necessary, RACS radio contours
were drawn interactively in {\sc Aladin} and overlaid on DES and/or
unWISE images. The unWISE images were especially helpful in suggesting
a likely host for sources with widely separated lobes without an
obvious radio nucleus and without a prominent galaxy or quasar 
near their geometrical center. Occasionally the unWISE images helped 
us to decide between two optical counterparts very close to each other.
In cases of lobes with very different
radio brightness, a higher probability was assigned to hosts nearer
to the brighter lobe (cf.~\cite{delarosa19}).  Whenever
the optical hosts were too faint or uncatalogued in DES\,DR1 we
consulted the Dark Energy Spectroscopic Instrument (DESI, \cite{dey19}) 
DR9 images at \url{https://www.legacysurvey.org} and usually found
the corresponding DESI object and its $r$-band magnitude. 
In case of doubt about 
the likely host we generally chose the brighter or lower-redshift host, 
such that the derived LLS should serve as a lower limit.

During this inspection we also measured the LAS for each accepted candidate,
and classified their radio morphology into one or more types as listed in 
Table~\ref{tab:morph}, appended by a ``?" symbol to indicate uncertainty.
We avoided any systematic overestimate of the LAS; e.g., for
FR\,II sources with bright hotspots we did not measure the LAS between 
opposite 3-$\sigma$
contours, but rather between the centres of the outer half-circles of these
contours of each hotspot. Only for rather faint or diffuse lobes we
measured out to about the 3-$\sigma$ contours (see \cite{kuzmicz21} sect.\ 2,
for a discussion of LAS measures). For bent-tailed sources we
measured the LAS along a straight line between the most separate
diametrically opposite emission regions of the source, and not along their
curved emission ridges.

\begin{table}[H]  
\centering
\caption{\small List of radio-morphological types we assigned to our sources.}
{\small
\begin{tabular}{ll}
\toprule
\textbf{Morph.Type}          & \textbf{Description}\\
\midrule
FRI         & radio brightness is ``edge-darkened", fading away with distance from the host \\
FRII        & classical double with ``edge-brightened" outer lobes \\
FRI/II      & source shows characteristics of both FR\,I and FR\,II \\
FRIIncor  & widely separated double with no obvious radio core at the optical host \\
FRIIcoredom & radio nucleus very strong compared to the lobes \\
FRIInaked   & no evidence for radio trails or bridges: lobes are unresolved hotspots \\
FRIIplume(s) & diffuse emission regions displaced sideways from source major axis  \\
FRIIrelic   & lobes slightly diluted \\
FRIIremn & lobes very diluted/inflated and of low surface brightness \\
DDRG        & double-double or ``restarted" radio galaxy: an inner and outer pair of lobes \\
hymor       & hybrid morphology: one arm or lobe is of type FR\,I, the other type FR\,II \\
WAT         & wide-angle tailed RG with outer lobes bent in the same direction, C-shaped) \\
NAT         & narrow-angle tailed RG: host galaxy is located at one end of radio emission \\
precess     & Z- or S-symmetry, suggesting precession of the radio jet axis  \\
asym        & length or flux ratio of opposite lobes is near two or more \\
\bottomrule
\end{tabular}}
\label{tab:morph}
\end{table}

Often the likely optical hosts, or objects superposed on the suspected
lobes, appeared stellar, either by visual impression or confirmed by
the stellarity index provided in the DES\,DR1 catalogue. To distinguish
between stars and quasars, we made use of both the measurements of
parallax and proper motion in the Global Astrometric Interferometer 
for Astrophysics (Gaia) early data release~3 (Gaia\,EDR3, \cite{gaiaEDR3})
which should be consistent with zero for quasars, as well as their 
$W1-W2$ and $W2-W3$ colors in the AllWISE catalogue \cite{cutri13, nikutta14}.
Based on this information we discarded the likely stars as possible radio 
galaxy hosts, or else labelled the host as quasar candidate or ``Qc'' in 
Table~\ref{tab:grglist}. We also took note of any optical peculiarities 
of the host like stellar/QSO or spiral morphology, presence of optical
shells, interactions or perturbations, etc.

Obvious relic-type radio sources in clusters of galaxies,
not associated with any single cluster galaxy, as well as nearby
late-type spiral galaxies were discarded as radio galaxy candidates.
We maintained a few possible candidates for so-called spiral double
radio galaxy AGN (SDRAGN, \cite{mao15}), though all of these had
linear sizes $\ll$1\,Mpc.

Scrutinizing these $\sim$1330 candidates required $\sim40$\,net hours, and led 
us to discard $\sim$30\% of them,
resulting in a list of precise host positions, LAS measures, and a crude 
radio-morphological classification for $\sim$1000 extended radio sources,
in addition to the $\sim$260 objects initially suspected on SUMSS images and
confirmed by us in RACS.

\subsection{Culling Complementary Data for the Radio Galaxy Hosts}

We first queried the positions of the $\sim$1300 newly discovered 
ERG hosts from RACS in the NASA/IPAC Extragalactic Database (NED,
\url{ned.ipac.caltech.edu}) to find spectroscopic redshifts
for 68 of them, mostly among the optically brightest hosts. We then 
used the VizieR service at the Strasbourg astronomical Data Center
(CDS, \cite{ochsenbein00}) at the URL
\url{https://vizier.u-strasbg.fr/viz-bin/VizieR}
to search for host names and complementary data for these hosts.
For all hosts we searched names (if available) from the 2MASX
catalogue \cite{2masx} (CDS VII/233), their exact positions,
names and $r$-band magnitudes, as well as stellarity index from
DES\,DR1 \cite{abbott18} (CDS I/357), and the WISEA names
and magnitudes in all four WISE bands from AllWISE \cite{cutri13}
(CDS II/328), as well as W1 and W2 magnitudes from the deeper
CatWISE2020 \cite{marocco21} (CDS II/365).

\subsection{Photometric Redshifts and Conversion from Angular to Linear Size} \label{sect:zphot}

We used several different sources outside VizieR for photometric redshifts ($z_{phot}$). For 2MASX galaxies we searched the 2MASS Photometric Redshift catalogue
(2MPZ, \cite{bilicki14}), and the hosts found in AllWISE \cite{cutri13}
were searched in the WISE-Supercosmos photometric redshift catalogue  
\cite{bilicki16} offering 20.4 and 57.9 million objects in the ``main'' and 
``reject'' catalogues, respectively. Previous experience showed that
the $z_{phot}$ from the ``reject'' catalogue are mostly sound and were used 
as well.

The $z_{phot}$ values offered by DESI~DR9 \cite{zhou21} were extracted from the 
DESI Legacy Survey DR9 through the Notebook Server of the NOIRLab Data Lab 
Query Interface at \url{https://datalab.noirlab.edu/query.php}. Since 
DESI\,DR9 data are stored in different tables, we cross-matched 
the main photometry table (\texttt{ls\_dr9.tractor}) and the one
with photometric redshifts (\texttt{ls\_dr9.photo\_z}) via the \texttt{objectID}
searchig a radius of $~2''$ from the host position, and requesting object
coordinates, $g,r,z$-magnitudes and median and mean $z_{phot}$.
For a few  objects the $z_{phot}$ values from DESI~DR9 were obtained 
directly at \url{viewer.legacysurvey.org}.

The DESI\,DR9 $z_{phot}$ catalogue also lists the standard deviation (which
we call $\Delta z$ here) of the the $z_{phot}$ probability distribution function.
We found the latter to be $\Delta z\sim$0.01 below $z_{phot}\sim$0.4,
gradually rising to $\Delta z\sim$0.05 near $z_{phot}\sim$0.8 and
to $\Delta z\sim$0.15 for $z_{phot}\gtrsim$1.0. For 23 (13\%) of our new 178
GRS in Table~\ref{tab:grglist} the $\Delta z$ values exceed 0.17 and 
reach up to 0.6, but these objects
are predominantly QSO candidates (for which we use additional $z_{phot}$
estimates, see below) and some very faint galaxies.
For 41 GRS up to a redshift of $\sim$0.6 we had $z_{phot}$ values from
both DESI\,DR9 and \cite{bilicki16}. The difference (DESI\,DR9 minus
\cite{bilicki16}) has a mean of 0.03 and a standard deviation of 0.09.
Averaging the latter values should further reduce their uncertainty.
Although the $z_{phot}$ values are considerably more uncertain at higher
redshift, the LLS is much less sentitive to the redshift for a given
LAS at redshifts between 1 and 2 (see our Fig.~6).

It is known that for QSOs it is notoriously difficult to estimate
redshifts, and a large fraction of starlike QSO candidates among our
hosts have clearly underestimated $z_{phot}$ values in DES\,DR1.
For some of them we found estimated redshifts in \cite{flesch21}, and,
in addition, for all our quasar candidates we use MIR colors 
$W1-W2$ and $W2-W3$ from AllWISE where available and consulted Fig.~2 of 
\cite{krogager18} to estimate their redshift. We combined all the
redshifts found, and adopted a reasonable average of these if more than
one was available, and provide these in Table~\ref{tab:grglist}.

In order to find the physically largest objects in our sample, we 
converted the sources' LAS to their LLS, using a script adapted
from the Cosmology Calculator (\cite{wright06},  \url{http://www.astro.ucla.edu/~wright/CosmoCalc.html}), and
based on the adopted redshift ($z_{spec}$ or $z_{phot}$) and
cosmological parameters as listed at the end of Sect.~\ref{intro}.
We stress that the
LLS is always a lower limit to the 3-D extent of the source and
we make no inference about its orientation with respect to the
line of sight as done e.g.\ in \cite{machalski21}.

\subsection{Determination of Radio Flux Densities and Luminosities} \label{sect:fluxint}

Cutouts of at least 2$\times$LAS on a side  were obtained for each 
source from both RACS and CRACS using a script as suggested at
\url{http://alasky.u-strasbg.fr/hips-image-services/hips2fits}.
These were then passed through the \texttt{breizorro} masking
program (see \url{https://github.com/ratt-ru/breizorro}) which creates
a binarized image assigning a value of~1 to every pixel brighter than
n\,$\sigma$, where $\sigma$ is the noise level, and values of~0 
elsewhere. By trial and error we found that a 3-$\sigma$
threshold worked best.

These binarized images were then run through the \texttt{python
matplotlib} program which calculates contours around the areas assigned
a value of~1, and then colors the boundaries of the ten contours
containing the largest areas. These are usually the ones containing the
real source confines, plus occasional unrelated sources, but sometimes 
an isolated central nuclear source
could have a contour area smaller than the ten largest.

The contour areas considered to contain the emission from the 
extended radio source are then selected manually (by clicking the
mouse on these) to be run through 
the source parameter analysis. Usually, we selected between two such areas
(e.g.\ for a source with only two lobes and no radio core) and 
up to about five such areas (e.g.\ for a source with a radio core
and two separated emission areas for each lobe on opposite sides
of the host), but occasionally the presence of a radio core and
emission bridge connecting the lobes leads to the source being
contained in a single such area.

The pixels making up the individual contour boundaries, along with
the x,y positions inside contours that were selected in the previous
operation are then parsed to the script that calculates source
parameters. It first multiplies the original image
by the \texttt{breizorro} mask. For a selected contour it then sums
up the brightness values of all pixels inside a contour and finally
normalizes by the number of pixels per beam solid angle to obtain the
flux density within a particular contour. To figure out which pixels
lie inside a contour the \texttt{python} package \texttt{Shapely}
(see \url{https://github.com/Toblerity/Shapely}) was used. A contour 
equates to a polygon in \texttt{Shapely} which can generate a MultiPolygon
from a collection of polygons.  \texttt{Shapely} is also used to find
out which contours are to be used in the analysis by determining which
contours contain the x,y positions determined by the mouse clicks in
the previous step.  This allows one to calculate total flux densities,
largest angular size, etc., which in fact we use to estimate magnetic
field strength and energy densities for the individual lobes based on the
assumption of equipartition between particle and magnetic field energies
(see \cite{miley80}).  However, for the present paper we limit ourselves
to report only the total flux of each source, as well as the position 
angle of the radio source's major axis.

As a final step we compared the total flux densities obtained from
RACS and CRACS to select the most adequate total flux, avoiding as much as 
possible the contributions of unrelated sources, and these are
listed as $S_{888}$ in Table~\ref{tab:grglist}.
Rest-frame spectral radio powers at 888\,MHz were then obtained via
 $P_{888} = 4\pi~D_L^2(z)~S_{888}~(1+z)^{-(1+\alpha)}$, 
where $D_L(z)$ is the luminosity distance for the adopted cosmology and 
$(1+z)^{-(1+\alpha)}$ is the $k$-correction to convert from observed
to rest frequency and allows for the stretching of the spectrum with 
respect to the receiver bandwidth. In the absence of measured spectral
indices we assume $\alpha=-0.8$, 
which is slightly steeper than the average spectrum of radio galaxies 
of $\alpha=-0.7$ \cite{jackson01}, given that we are dealing mostly 
with objects dominated by emission from diffuse radio lobes.
We list the decimal logarithm of $P_{888}$ in Table~\ref{tab:grglist}.

\section{Results}   \label{results}

Having obtained the LLS for all $\sim$1440 accepted GRS
in our search area, we ranked them in decreasing order
of LLS and found $\sim$210 to exceed a size of 1\,Mpc. For these,
overlays of radio contours on DES $g,r,i-$band composites were
prepared and scrutinized again. As a consequence,  about 30 of these
were discarded for being likely separate sources, and 178 objects
were accepted as new GRS, of which we mark 14 as likely candidates
in Table~\ref{tab:grglist} due to uncertainties in either
their LAS, $z_{phot}$, or their host ID.

As a sideproduct our list of smaller radio galaxies contains
$\sim$300 ERGs with LLS=0.7--1.0\,Mpc, plus $\sim$400 
with LLS=0.5--0.7\,Mpc, and 550 smaller ones. All these
require further scrutinizing and will be published elsewhere. 
However, these numbers serve to
demonstrate the significant effort necessary to find physically
large sources among objects selected purely by their angular size.
For example, of the 558 surviving radio galaxies with LAS$>2.0'$,
only 32\% turned out to be GRS larger than 1\,Mpc, and of all
the 1038 with LAS$>1.4'$, only 46\% turned out to be ERGs
with LLS$>0.7$\,Mpc.  Given that from the candidates originally 
selected on the RACS images, about 30\% had to be discarded upon 
closer inspection, this implies that to find a single GRS larger
than 1\,Mpc one has to inspect on average 4.4 candidates larger
than 2.0$'$ and to find a single ERG larger than 0.7\,Mpc one
has to inspect on average 3 candidates larger than 1.4$'$.

\subsection{The List of 178 new and 3 known GRS in the Search Area}   \label{sect:grglist}

\begin{figure}[H]	
\includegraphics[width=13.5cm]{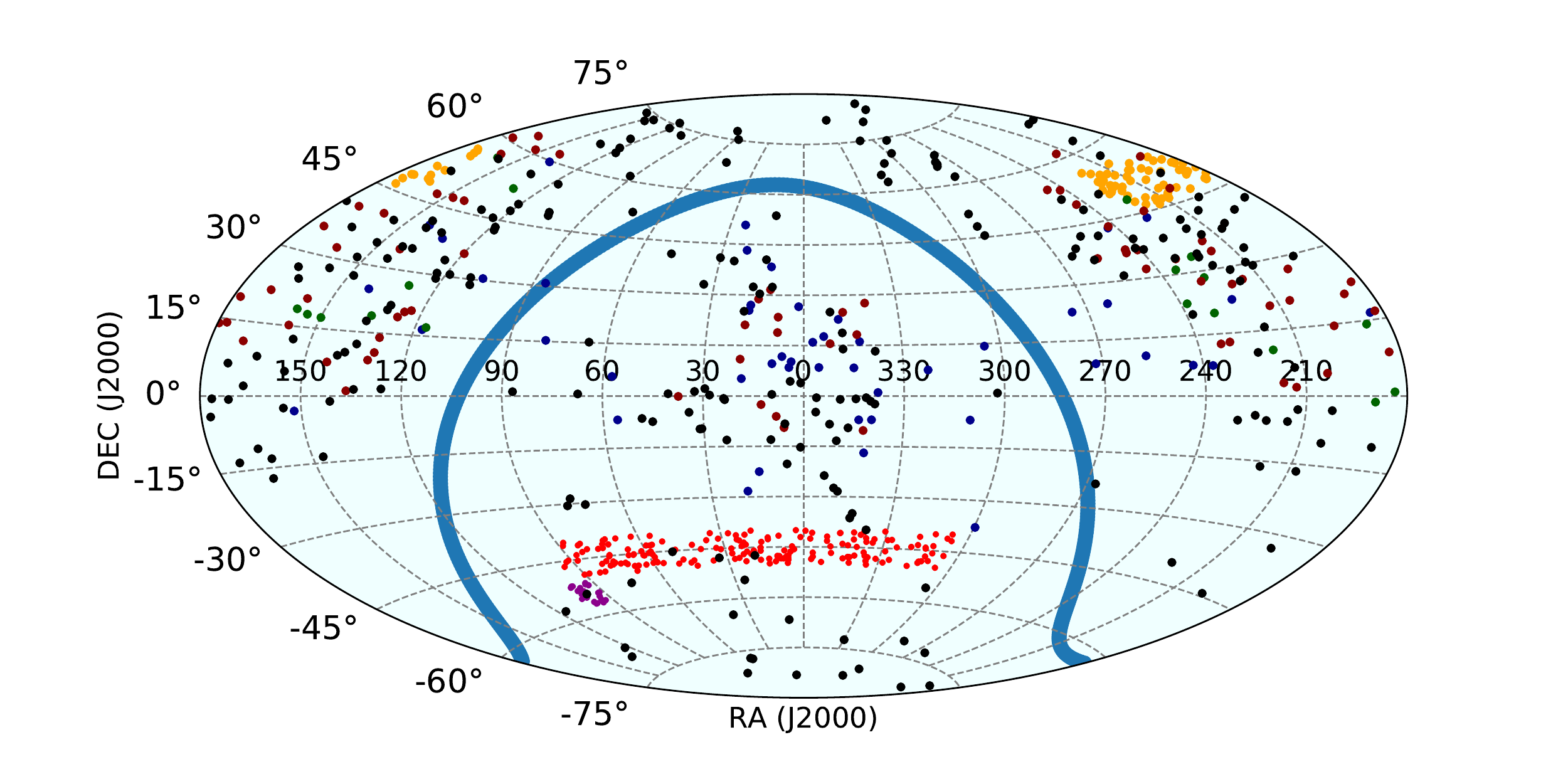}
\caption{All-sky Aitoff projection of the distribution of our newly discovered GRS
in red, together with the GRS from six major published GRS lists, namely in black 
from \cite{kuzmicz18} in orange from \cite{dabhade20a}, in dark blue 
from \cite{dabhade20b}, in dark green from \cite{koziel20}, in dark magenta 
from \cite{brueggen21}, and in dark red from \cite{kuzmicz21}. The thick blue line shows the area within $|b|\le2^{\circ}$ from the Galactic plane.}
\label{fig:skydist}
\end{figure}

In what follows we shall limit ourselves to the sample of 178 new
and 3 previously published GRS larger than 1\,Mpc. Their sky distribution
is plotted in Figure~\ref{fig:skydist}. In about
20 cases the host could not be determined with certainty, and either
the brightest in optical or MIR was chosen, such that its redshift
and linear size can be considered a lower limit. The basic properties
of these 181 GRS are listed in Table~\ref{tab:grglist}, the columns
of which are as follows: 
(1) name of the GRS derived from truncation of the sexagesimal
equatorial J2000 coordinates of the optical host, appended with a
``C'' if the GRS is considered a candidate; for each object
we indicate in which survey it was noticed first: P = previously
published (3 objects), A = ATESP (3 objects), E\,=\,EMU Pilot Survey
(8 objects), S = SUMSS (45 objects), R = RACS (122 objects);
(2,3) RA and Dec (J2000) of the optical host in decimal degrees;
(4) the largest angular size of the radio emission;
(5) the adopted redshift and type: s for spectroscopic, p for
photometric, and e for estimated, with question marks indicating uncertainty 
due to inconsistent $z_{phot}$ values found, see also Sect.~\ref{sect:zphot};
(6) references used to select the adopted redshift:
1  \cite{zhou21},     
2  \cite{bilicki16},  
3  \cite{bilicki14},  
4  \cite{burgess06},  
5  \cite{colless01},  
6  \cite{jones92},    
7  \cite{jones09},    
8  \cite{krogager18}, 
9  \cite{flesch21}    
10 \cite{danziger78}, 
11 \cite{danziger83}  
12 \cite{loveday96},  
13 \cite{wisotzki00}  
14 \cite{shu19};       
(7) largest linear size of the radio emission;
(8,9) name and type of the host object, where G=galaxy, GP=galaxy pair,
GQ=galaxy or QSO, Q=QSO, Qc=QSO candidate, with
a ``?'' sign indicating uncertainty;
(10) $r$-band magnitude from DES;
(11) integrated flux density from RACS or CRACS (see Sect.~\ref{sect:fluxint});
(12) decimal logarithm of spectral radio power at 888\,MHz;
(13) position angle of the major axis of the radio source,
measured from North through East;
(14) radio morphology according to Table~\ref{tab:morph}.

Host redshifts range from 0.02 to $\sim$2.0, but only 10 of the 178
new GRS have spectroscopic redshifts. For the 146 host galaxies the
median $r$-band magnitude and redshift are 20.9 and 0.64, while for
the 32 QSOs or candidates these are 19.7 and 0.75. The lowest total radio
flux densities are 5.0\,mJy with 10 objects fainter than 10\,mJy. The
lobe brightness of various GRS barely exceeds 1\,mJy\,beam$^{-1}$.
We compare the properties of our new sample with those of
six previous large compilations of GRS in Sect.~\ref{sect:compar}.

\subsection{The Variety of Source Morphologies, Sizes and Radio Luminosities}

Within the classification scheme we used (see Table~\ref{tab:morph})
we find that the vast majority of the newly found GRS are FR\,II (166 or 93\%),
of which 30 do not have a radio core detected (i.e.\ S$_{888,core}\lesssim0.5$\,mJy),
four have virtually unresolved lobes (tagged as ``naked" 
\\
\startlandscape
 \begin{specialtable}[H]
 \widetable
\caption{List of 181 giant radio sources in our search area (see text for column descriptions). The table will also be available at http://cdsarc.u-strasbg.fr/viz-bin/cat/J/other/Galax/9.xx} \label{tab:grglist}
  \begin{tabular}{lcclllclllrcrl}

 ine
    ~~~~(1)          &    (2)       &   (3)         &   ~(4)    &    ~(5)  &   (6)  &     (7)  &    ~~~~~~~~~(8)    &   (9)   & (10)  &   (11)  &   (12)   &   (13)    &  (14)  \\
  Name, origin   &   RA$_J$     & Dec$_J$     &   LAS    & ~z, ztype & Ref(z) &  LLS     & ~~~~~~Hostname  & Host    & rmag   & S$_{888}$~ & log\,P$_{888}$ & RPA  & Radio\,Morphology \\
                & ($^{\circ}$) &  ($^{\circ}$) &  ~($'$)   &         &        &  (Mpc)   &           &  Type    & (mag)  &  (mJy)  &    (W/Hz)     &   ($^{\circ}$) &        \\
\midrule
J0004--4205~~~~~S & 001.1349 & --42.0852 & ~~2.95  & ~~~0.53~~~~~p & 1,2      & ~~~1.11 & DES~J000432.36--420506.7~~ & G  & 20.80   &   84.~~~& 25.93 &  27 & FRIIrelic,ncor       \\
J0011--4001~~~~~A & 002.7774 & --40.0253 & ~~2.6   & ~~~0.881~~~s  & 13       & ~~~1.08 & HE~0008--4018~~~~~~~~~~~~~ & Q  & 17.44   &    30.3 & 26.01 &  59 & FRIIcoredom          \\
J0014--4542C~~R   & 003.6338 & --45.7011 & ~~4.2   & ~~~0.296~~~p  & 1,3      & ~~~1.11 & DES~J001432.11--454204.1~? & G  & 19.53   &    13.4 & 24.55 &  56 & FRIIremn,ncor,asym   \\
J0016--4541~~~~~R & 004.0419 & --45.6857 & ~~2.2   & ~~~1.07~~~~~p & 1,8      & ~~~1.07 & DES~J001610.05--454108.6~~ & G  & 22.49   &    13.9 & 25.87 & 112 & FRIIremn             \\
J0017--4449~~~~~R & 004.3794 & --44.8306 & ~~2.05  & ~~~1.12~~~~~p & 1        & ~~~1.01 & DES~J001731.05--444950.2~~ & G  & 22.95   &    61.9 & 26.57 & 141 & FRIIrelic            \\
J0020--4625~~~~~R & 005.0739 & --46.4173 & ~~2.44  & ~~~0.65~~~~~p & 1,2      & ~~~1.01 & DES~J002017.74--462502.2~~ & G  & 20.66   &    87.4 & 26.16 &  12 & FRIIremn             \\
J0023--4732~~~~~R & 005.9779 & --47.5410 & ~~3.25  & ~~~0.52~~~~~p & 1        & ~~~1.21 & DES~J002354.69--473227.6~? & G  & 21.49   &    32.4 & 25.50 &  22 & FRIIncor,naked       \\
J0024--4816~~~~~R & 006.0832 & --48.2691 & ~~5.8   & ~~~0.268~~~p  & 1,2      & ~~~1.43 & DES~J002419.94--481608.5~~ & G  & 18.37   &    34.8 & 24.87 &  50 & FRIIrelic,ncor       \\
J0024--4848~~~~~R & 006.0838 & --48.8115 & ~~2.37  & ~~~0.90~~~~~p & 1,8      & ~~~1.11 & DES~J002420.12--484841.5~~ & GQ & 22.65   &    31.0 & 26.04 & 138 & FRII                 \\
J0025--4946~~~~~R & 006.3780 & --49.7787 & ~~3.42  & ~~~0.51~~~~~p & 1        & ~~~1.27 & DES~J002530.72--494643.4~~ & G  & 20.70   &    12.9 & 25.08 &  11 & FRIIcoredom          \\
J0027--4746~~~~~S & 006.7520 & --47.7750 & ~~4.47  & ~~~0.84~~~~~p & 1        & ~~~2.05 & DES~J002700.50--474630.0~~ & G  & 21.42   &   211.9 & 26.80 & 149 & FRII                 \\
J0029--4601~~~~~R & 007.4194 & --46.0246 & ~~2.79  & ~~~0.50~~~~~p & 1        & ~~~1.02 & DES~J002940.66--460128.7~~ & G  & 20.26   &     7.2 & 24.80 & 152 & FRI/II,faint         \\
J0031--4826~~~~~R & 007.9993 & --48.4343 & ~~3.09  & ~~~0.65~~~~~p & 1        & ~~~1.28 & DES~J003159.82--482603.5~~ & G  & 21.59   &    42.1 & 25.84 & 167 & FRIIncor             \\
J0035--4824~~~~~R & 008.9547 & --48.4161 & ~~2.9   & ~~~0.94~~~~~p & 1        & ~~~1.37 & DES~J003549.12--482458.0~~ & G  & 23.02   &    59.8 & 26.37 &  80 & FRIIncor ?           \\
J0036--4138~~~~~R & 009.1483 & --41.6435 & ~~2.55  & ~~~0.57~~~~~p & 1        & ~~~1.00 & DES~J003635.58--413836.7~~ & G  & 21.35   &    27.6 & 25.52 & 102 & FRIIremn,ncor        \\
J0038--4741C~~R   & 009.5398 & --47.6938 & ~~2.3   & ~~~1.17~~~~~p & 1        & ~~~1.14 & DES~J003809.54--474137.8~? & G  & 23.49   &    37.9 & 26.40 & 160 & FRIIncor             \\
J0040--4817~~~~~R & 010.1491 & --48.2885 & ~~7.08  & ~~~0.53~~~~~p & 1,2      & ~~~2.67 & DES~J004035.77--481718.6~~ & G  & 20.22   &    14.8 & 25.18 & 177 & FRI/II               \\
J0041--4738~~~~~R & 010.3071 & --47.6373 & ~~3.84  & $>$0.4~~~~?~p & 1,3,8,9  & $>$1.24 & DES~J004113.71--473814.4~~ & Qc & 18.10   &     5.0 & 24.42 & 139 & FRII                 \\
J0044--4915~~~~~S & 011.1276 & --49.2545 & ~~2.4   & ~~~0.7~~~~?~p & 1,8      & ~~~1.03 & DES~J004430.62--491516.1~~ & Qc & 17.00   &   267.1 & 26.72 & 120 & FRII                 \\
J0052--4238~~~~~R & 013.1770 & --42.6461 & ~~2.4   & ~~~0.87~~?\,p & 1,8,9    & ~~~1.11 & DES~J005242.47--423845.8~~ & Qc & 19.97   &    14.7 & 25.68 &  12 & FRIIremn             \\
J0052--4206~~~~~R & 013.1866 & --42.1058 & ~~2.14  & ~~~0.92~~~~~p & 1        & ~~~1.01 & DES~J005244.79--420620.8~~ & G  & 22.79   &    19.0 & 25.85 &   0 & FRIIrelic,ncor       \\
J0054--4825~~~~~R & 013.6751 & --48.4269 & ~~2.55  & ~~~0.80~~~~~p & 1        & ~~~1.15 & DES~J005442.02--482536.7~~ & G  & 22.26   &    37.7 & 26.00 &  82 & FRIIremn             \\
J0054--4952~~~~~R & 013.7436 & --49.8739 & ~~9.67  & ~~~0.47~~~~~p & 1,2      & ~~~3.42 & DES~J005458.45--495226.0~~ & G  & 19.74   &    24.0 & 25.26 & 156 & FRII                 \\
J0103--4324~~~~~R & 015.7548 & --43.4082 & ~~2.8   & ~~~0.874~~~p  & 1        & ~~~1.30 & DES~J010301.15--432429.5~~ & G  & 21.51   &    17.8 & 25.77 & 131 & FRIIrelic            \\
J0105--4505~~~~~S & 016.3425 & --45.0881 & ~~2.85  & ~~~0.69~~~~~p & 1,4      & ~~~1.21 & DES~J010522.19--450517.2~~ & G  & 20.23   &  4291.9 & 27.91 &  56 & FRII                 \\
J0106--4602~~~~~S & 016.5371 & --46.0387 & ~~3.39  & ~~~0.478~~~p  & 1        & ~~~1.21 & DES~J010608.90--460219.3~~ & G  & 20.20   &    37.0 & 25.47 &  30 & FRIIncor             \\
J0108--4843~~~~~R & 017.0196 & --48.7175 & ~~4.2   & ~~~0.503~~~p  & 1        & ~~~1.54 & DES~J010804.70--484302.9~? & G  & 20.65   &     7.5 & 24.83 & 118 & FRIIremn             \\
J0116--4000~~~~~A & 019.0221 & --40.0130 & ~~3.42  & ~~~1.06~~~~~p & 1        & ~~~1.66 & DES~J011605.29--400046.6~~ & G  & 22.95   &    59.8 & 26.49 &  27 & FRII                 \\
J0116--4722~~~~~P & 019.1043 & --47.3782 &  11.2   & ~~~0.146~~~s  & 11       & ~~~1.70 & 2MASX~J01162507--4722406~~ & G  & 15.62   & 4210.~~~& 26.37 & 169 & FRII,DDRG            \\
J0116--4555~~~~~R & 019.1479 & --45.9257 & ~~4.5   & ~~~1.015~~~p  & 1        & ~~~2.17 & DES~J011635.50--455532.6~~ & Qc & 23.72   &    15.3 & 25.86 & 132 & FRIInaked            \\
J0121--4656~~~~~R & 020.3779 & --46.9399 & ~~2.68  & ~~~0.813~~~p  & 1        & ~~~1.21 & DES~J012130.69--465623.6~~ & G  & 22.29   &    39.8 & 26.04 &  27 & FRIIncor             \\
J0125--4110~~~~~R & 021.4841 & --41.1696 & ~~8.0~  & ~~~0.11852\,s & 7        & ~~~1.03 & 2MASX~J01255616--4110102~~ & G  & 15.93   &    90.1 & 24.51 &  27 & FRI/II,WAT?          \\
J0126--4407~~~~~R & 021.7427 & --44.1182 & ~~3.9~  & ~~~0.364~~~p  & 1,2      & ~~~1.19 & DES~J012658.24--440705.6~~ & G  & 19.18   &    18.9 & 24.91 & 176 & FRI/IIremn           \\
\bottomrule
\end{tabular}   \end{specialtable}
\begin{specialtable}[H]\ContinuedFloat
\widetable\small
\caption{{\em Cont.}}
\begin{tabular}{lcclllclllrcrl}

 ine
    ~~~~(1)          &    (2)       &   (3)         &   ~(4)    &    ~(5)  &   (6)  &     (7)  &    ~~~~~~~~~(8)    &   (9)   & (10)  &   (11)  &   (12)   &   (13)    &  (14)  \\
  Name, origin   &   RA$_J$     & Dec$_J$     &   LAS    & ~z, ztype & Ref(z) &  LLS     & ~~~~~~Hostname  & Host    & rmag   & S$_{888}$~ & log\,P$_{888}$ & RPA  & Radio\,Morphology \\
                & ($^{\circ}$) &  ($^{\circ}$) &  ~($'$)   &         &        &  (Mpc)   &           &  Type    & (mag)  &  (mJy)  &    (W/Hz)     &   ($^{\circ}$) &        \\
\midrule
J0127--4610~~~~~R & 021.8368 & --46.1770 & ~~4.67  & ~~~0.72~~~~~p & 1,9      & ~~~2.03 & DES~J012720.83--461037.0~~ & Qc & 19.78   &   112.2 & 26.37 & 174 & FRIIasym             \\
J0129--4330~~~~~R & 022.3443 & --43.5151 & ~~2.2~  & ~~~0.919~~~p  & 1        & ~~~1.03 & DES~J012922.62--433054.3~~ & G  & 24.32   &    11.3 & 25.62 & 164 & FRIIncor             \\
J0131--4901~~~~~R & 022.9477 & --49.0263 & ~~2.43  & ~~~1.0~~~~~?~e & -       & ~~~1.17 & (DESI~J022.9477--49.0263)~ & G  & 24.5\,? &    45.4 & 26.32 &  17 & FRIIremn             \\
J0133--4655~~~~~R & 023.3367 & --46.9171 & ~~2.5~  & ~~~0.65~~~~~p & 1,9      & ~~~1.04 & DES~J013320.80--465501.4~~ & G  & 20.25   &    38.4 & 25.80 &  10 & FRIIremn             \\
J0133--4431~~~~~R & 023.4138 & --44.5170 & ~~2.1~  & ~~~1.048~~~p  & 1        & ~~~1.02 & DES~J013339.30--443101.1~~ & G  & 22.34   &   274.1 & 27.14 &   6 & FRII                 \\
J0135--4249~~~~~R & 023.8273 & --42.8276 & ~~2.64  & ~~~2.0~~~~?~p & 8        & ~~~1.33 & DES~J013518.56--424939.4~~ & Qc & 22.67   &    34.7 & 26.91 & 168 & FRII                 \\
J0138--4116~~~~~R & 024.6351 & --41.2714 & ~~2.75? & ~~~0.954~~~p  & 1        & ~~~1.31 & DES~J013832.41--411616.9~~ & G  & 22.03   &    12.0 & 25.69 & 172 & FRI/II,hymor?        \\
J0138--4231~~~~~S & 024.6632 & --42.5272 &  46.~~  & ~~~0.02123\,s & 5        & ~~~1.19 & NGC~641; ESO 244--G042     & G  & 12.04   &   334.6 & 23.53 &  37 & FRI/II               \\
J0141--4759~~~~~R & 025.3820 & --47.9861 & ~~3.0~  & ~~~0.984~~~p  & 1        & ~~~1.44 & DES~J014131.67--475909.9~~ & G  & 22.10   &   263.6 & 27.06 & 139 & FRII                 \\
J0148--4814C~~R   & 027.2305 & --48.2365 & ~~4.2~  & ~~~0.264~~~p  & 1,2      & ~~~1.03 & DES~J014855.31--481411.2~~ & G  & 18.44   &    24.0 & 24.69 & 114 & FRIIremn             \\
J0149--4036~~~~~S & 027.3344 & --40.6159 & ~~2.4~  & ~~~0.7~~~~?~p & 1,8,9    & ~~~1.03 & DES~J014920.26--403657.3~~ & Qc & 19.33   &    46.2 & 25.95 &  35 & FRI/IIremn           \\
J0150--4507~~~~~S & 027.5045 & --45.1258 & ~~6.64  & ~~~0.25~~~~~p & 1,2      & ~~~1.56 & 2MASX~J01500108--4507331~~ & G  & 17.59   &   105.7 & 25.28 &  10 & FRII                 \\
J0150--4634~~~~~S & 027.6889 & --46.5833 & ~~2.48  & ~~~0.88~~~~~p & 1        & ~~~1.15 & DES~J015045.33--463459.8~~ & G  & 22.14   &    33.3 & 26.05 &  21 & FRIIremn,precess?    \\
J0205--4206~~~~~S & 031.4603 & --42.1121 & ~~4.6~  & ~~~0.31941\,s & 7        & ~~~1.28 & HE~0203--4221              & Q  & 15.81   &   261.8 & 25.92 &  68 & FRIIrelic            \\
J0206--4514~~~~~R & 031.5664 & --45.2382 & ~~3.25  & ~~~0.631~~~p  & 1        & ~~~1.33 & DES~J020615.92--451417.3~~ & G  & 20.47   &    61.1 & 25.97 &  55 & FRIIncor             \\
J0212--4447~~~~~R & 033.1357 & --44.7957 & ~~3.38  & ~~~1.191~~~p  & 1        & ~~~1.68 & DES~J021232.56--444744.4~~ & Qc & 23.69   &    20.6 & 26.15 & 144 & FRIInaked            \\
J0213--4744~~~~~P & 033.2902 & --47.7370 & ~~6.65  & ~~~0.220~~~s & 10        & ~~~1.42 & 2MASX~J02130961--4744128~~ & G  & 17.35   & 2050.~~~& 26.45 & 171 & FRIIplumes           \\
J0215--4024~~~~~S & 033.8490 & --40.4069 & ~~3.16  & ~~~0.42~~~~~p & 1,2      & ~~~1.05 & DES~J021523.75--402424.9~~ & G  & 20.34   &   585.5 & 26.54 & 126 & FRIIplumes           \\
J0223--4826~~~~~S & 035.8849 & --48.4483 & ~~2.34  & ~~~0.70~~~~~p & 1        & ~~~1.00 & DES~J022332.36--482654.0~~ & G  & 21.03   &   347.3 & 26.83 & 158 & FRII                 \\
J0223--4226~~~~~R & 035.9365 & --42.4352 & ~~4.0~  & $>$0.650~~~p  & 1        & $>$1.66 & DES~J022344.76--422606.7~? & G  & 22.60   &    40.2 & 25.82 & 176 & FRII                 \\
J0236--4456~~~~~R & 039.1024 & --44.9405 & ~~3.42  & ~~~0.859~~~p  & 1        & ~~~1.58 & DES~J023624.58--445625.6~~ & G  & 21.75   &    58.3 & 26.27 & 156 & FRII                 \\
J0247--4334~~~~~S & 041.8256 & --43.5719 & ~~3.65  & ~~~1.25~~~~~p & 1,8      & ~~~1.83 & DES~J024718.13--433418.9~~ & GQ & 22.42   &    52.3 & 26.61 &  26 & FRII                 \\
J0252--4941~~~~~R & 043.1323 & --49.6991 & ~~2.85  & ~~~0.664~~~p  & 1        & ~~~1.20 & DES~J025231.75--494156.7~~ & G  & 20.67   &    16.7 & 25.46 &   6 & FRIIrelic,coredom    \\
J0253--4806~~~~~R & 043.4708 & --48.1082 & ~~3.0~  & ~~~0.623~~~p  & 1        & ~~~1.22 & DES~J025352.99--480629.6~~ & G  & 20.37   &    15.9 & 25.37 &  31 & FRIIrelic            \\
J0259--4840~~~~~S & 044.9775 & --48.6795 & ~~3.22  & ~~~0.93~~~~~p & 1        & ~~~1.52 & DES~J025954.60--484046.2~~ & G  & 21.63   &   325.0 & 27.09 & 111 & FRII                 \\
J0310--4740~~~~~S & 047.5660 & --47.6774 & ~~3.38  & ~~~0.55~~~~~p & 1        & ~~~1.30 & DES~J031015.84--474038.5~? & G  & 20.21   &   204.4 & 26.35 & 106 & FRII                 \\
J0315--4443~~~~~R & 048.8579 & --44.7257 & ~~3.5~  & ~~~0.362~~~p  & 1,2      & ~~~1.06 & DES~J031525.88--444332.4~~ & G  & 19.17   &    21.8 & 24.96 & 106 & FRII                 \\
J0320--4845~~~~~R & 050.0403 & --48.7665 & ~~2.67  & ~~~0.645~~~p  & 1        & ~~~1.11 & DES~J032009.68--484559.2~? & Qc & 23.89   &    25.3 & 25.61 &  88 & FRIIncor             \\
J0320--4515~~~~~P & 050.2397 & --45.2530 & ~27.2~  & ~~~0.0633~s   & 6        & ~~~1.99 & ESO~248--G010; MSH 03@43~~ & G  & 14.33   & 5500.~~~& 25.72 &  47 & FRII                 \\
J0328--4208~~~~~R & 052.1877 & --42.1493 & ~~4.50  & ~~~0.55~~~~~p & 1,8,9    & ~~~1.73 & DES~J032845.03--420857.6~~ & Qc & 18.56   &    39.2 & 25.64 &  35 & FRIIasym             \\
J0342--4019~~~~~S & 055.6854 & --40.3272 & ~~4.5~  & ~~~0.69~~~~~p & 1        & ~~~1.92 & DES~J034244.49--401937.7~~ & G  & 20.45   &   382.8 & 26.86 & 107 & FRII                 \\
J0344--4817~~~~~R & 056.0404 & --48.2896 & ~~3.36  & ~~~0.380~~~p  & 1        & ~~~1.05 & DES~J034409.70--481722.4~~ & G  & 20.51   &    32.1 & 25.18 & 159 & FRIIncor             \\
J0346--4253~~~~~R & 056.5079 & --42.8945 & ~~2.75  & ~~~0.604~~~p  & 1        & ~~~1.11 & DES~J034601.89--425340.0~? & G  & 20.66   &    50.8 & 25.84 & 146 & FRIIrelic            \\
J0349--4302C~~R   & 057.3790 & --43.0424 & ~~2.7~  & ~~~0.879~~~p  & 1        & ~~~1.24 & DES~J034930.95--430232.4~? & G  & 23.56   &     6.7 & 25.35 &  31 & FRIIncor             \\
J0352--4626~~~~~R & 058.0616 & --46.4406 & ~~2.8~  & ~~~0.591~~~p  & 1        & ~~~1.12 & DES~J035214.77--462626.1~~ & G  & 20.89   &   125.7 & 26.22 &  82 & FRIIrelic            \\
J0352--4756~~~~~R & 058.1879 & --47.9477 & ~~2.7~  & ~~~0.927~~~p  & 1        & ~~~1.27 & DES~J035245.09--475651.8~~ & G  & 21.50   &     9.7 & 25.57 & 135 & FRIIrelic            \\
J0352--4536~~~~~R & 058.2474 & --45.6046 & ~~2.4~  & ~~~0.65~~~~~p & 1        & ~~~1.00 & DES~J035259.37--453616.6~~ & G  & 21.59   &    87.3 & 26.16 &  14 & FRIIrelic,asym       \\
\bottomrule
\end{tabular}   \end{specialtable}
\begin{specialtable}[H]\ContinuedFloat
\widetable\small
\caption{{\em Cont.}}
\begin{tabular}{lcclllclllrcrl}

 ine
    ~~~~(1)          &    (2)       &   (3)         &   ~(4)    &    ~(5)  &   (6)  &     (7)  &    ~~~~~~~~~(8)    &   (9)   & (10)  &   (11)  &   (12)   &   (13)    &  (14)  \\
  Name, origin   &   RA$_J$     & Dec$_J$     &   LAS    & ~z, ztype & Ref(z) &  LLS     & ~~~~~~Hostname  & Host    & rmag   & S$_{888}$~ & log\,P$_{888}$ & RPA  & Radio\,Morphology \\
                & ($^{\circ}$) &  ($^{\circ}$) &  ~($'$)   &         &        &  (Mpc)   &           &  Type    & (mag)  &  (mJy)  &    (W/Hz)     &   ($^{\circ}$) &        \\
\midrule
J0353--4446~~~~~R & 058.3342 & --44.7831 & ~~3.55  & ~~~0.353~~~s  & 13       & ~~~1.06 & HE~0351--4455              & Q  & 16.69   &   367.3 & 26.16 &  87 & FRIIbent             \\
J0354--4728~~~~~R & 058.5263 & --47.4737 & ~~2.95  & ~~~1.021~~~p  & 1        & ~~~1.42 & DES~J035406.30--472825.1~~ & G  & 22.90   &    90.0 & 26.63 &  60 & FRIIncor             \\
J0355--4258C~~R   & 058.7883 & --42.9702 & ~~2.28  & ~~~0.81~~~~~p & 1        & ~~~1.03 & DES~J035509.18--425812.7~? & G  & 22.19   &    48.1 & 26.12 & 119 & FRIIrelic            \\
J0356--4503~~~~~S & 059.1046 & --45.0665 & ~~2.1~  & ~~~1.13~~~~~p & 1        & ~~~1.03 & DES~J035625.11--450359.2~~ & G  & 24.12   &    69.9 & 26.63 &  80 & FRII                 \\
J0359--4800~~~~~R & 059.8806 & --48.0059 & ~~2.9~  & ~~~0.679~~~p  & 1        & ~~~1.23 & DES~J035931.34--480021.3~~ & G  & 20.73   &    28.2 & 25.71 &  10 & FRII                 \\
J0406--4544~~~~~S & 061.5326 & --45.7476 & ~~6.4~  & ~~~0.315~~~p  & 1,2      & ~~~1.77 & DES~J040607.82--454451.2~~ & G  & 18.26   &    77.0 & 25.37 & 131 & FRII                 \\
J0406--4429~~~~~S & 061.7500 & --44.4945 & ~11.8~  & ~~~0.1408~s   & 7        & ~~~1.76 & 2MASX~J04065999--4429400~~ & G  & 15.60   &    18.1 & 23.97 & 105 & FRI/IIremn           \\
J0407--4538~~~~~R & 061.8515 & --45.6480 & ~~2.23  & ~~~0.903~~~p  & 1        & ~~~1.04 & DES~J040724.34--453852.8~? & G  & 22.40   &    32.6 & 26.07 &  28 & FRIIrelic            \\
J0410--4012~~~~~R & 062.5550 & --40.2104 & ~~2.33  & ~~~0.91~~~~~p & 1        & ~~~1.09 & DES~J041013.20--401237.4~? & G  & 22.41   &    34.8 & 26.10 &  31 & FRIIbent             \\
J0412--4819C~~S   & 063.1106 & --48.3220 & ~~4.3~  & $>$0.26~~~~~p & 1        & $>$1.04 & DES~J041226.53--481919.2~~ & GP & 19.63   &    45.0 & 24.95 &  89 & FRIIncor ?           \\
J0420--4509~~~~~S & 065.2422 & --45.1645 & ~~7.2~  & ~~~0.32~~~~~p & 1,2,9    & ~~~2.00 & DES~J042058.13--450952.0~~ & G  & 19.04   &   440.6 & 26.14 & 133 & FRIIplume            \\
J0422--4518C~~R   & 065.5653 & --45.3096 & ~>7.5~? & ~~~0.388~~~p  & 1,2      & ~~~2.37 & DES~J042215.67--451834.6~~ & G  & 19.69   &    69.0 & 25.53 &  61 & FRI                  \\
J0425--4340C~~R   & 066.2899 & --43.6825 & ~~2.38  & ~~~0.895~~~p  & 1        & ~~~1.11 & DES~J042509.56--434056.8~~ & G  & 21.62   &    86.2 & 26.48 & 144 & FRII                 \\
J0425--4823~~~~~S & 066.3343 & --48.3871 & ~~2.46  & ~~~0.86~~~~~p & 1,8      & ~~~1.13 & DES~J042520.22--482313.5~~ & GQ & 21.09   &   137.0 & 26.64 & 129 & FRII                 \\
J0429--4517~~~~~S & 067.3708 & --45.2908 & ~~5.0~  & ~~~0.545~~~p  & 1,2      & ~~~1.91 & DES~J042928.98--451726.7~~ & G  & 19.74   &   195.5 & 26.33 &  15 & FRIIrelic            \\
J0431--4823~~~~~R & 067.9135 & --48.3954 & ~~2.76  & ~~~0.704~~~p  & 1        & ~~~1.19 & DES~J043139.24--482343.3~~ & G  & 21.71   &    51.9 & 26.01 & 122 & FRII                 \\
J0433--4948~~~~~R & 068.3529 & --49.8026 & ~~2.51  & ~~~0.592~~~p  & 1        & ~~~1.00 & DES~J043324.68--494809.2~? & G  & 21.46   &    27.2 & 25.55 &  35 & FRI/IIcoredom,hymor? \\
J0433--4022~~~~~R & 068.3570 & --40.3824 & ~~2.85  & ~~~0.555~~~p  & 1        & ~~~1.10 & DES~J043325.68--402256.7~~ & G  & 21.14   &    20.4 & 25.36 &   7 & FRIIrelic            \\
J0440--4742~~~~~R & 070.1712 & --47.7097 & ~~6.63  & ~~~0.302~~~p  & 1,2      & ~~~1.78 & DES~J044041.09--474234.9~~ & G  & 18.56   &    36.3 & 25.00 &  52 & FRIIremn             \\
J0442--4716~~~~~R & 070.5712 & --47.2812 & ~~3.4~  & ~~~0.356~~~p  & 1,2      & ~~~1.02 & DES~J044217.08--471652.4~~ & G  & 18.35   &     8.2 & 24.52 &  72 & FRIIrelic            \\
J0448--4151~~~~~S & 072.2195 & --41.8611 & ~~4.08  & ~~~0.694~~~p  & 1        & ~~~1.74 & DES~J044852.67--415140.0~~ & G  & 21.45   &    93.8 & 26.25 & 174 & FRII                 \\
J0449--4022~~~~~S & 072.2808 & --40.3793 & ~~2.2~  & ~~~0.9~~~~?~p & 1,8,9    & ~~~1.03 & DES~J044907.38--402245.4~~ & Qc & 19.65   &    35.5 & 26.10 & 159 & FRII                 \\
J0450--4722~~~~~S & 072.5735 & --47.3736 & ~~2.35  & ~~~1.5~~~~~~~p & 8,9     & ~~~1.19 & DES~J045017.64--472225.1~~ & Qc & 17.23   &   597.9 & 27.85 &  45 & FRII                 \\
J0452--4040~~~~~S & 073.2427 & --40.6699 & ~~4.05  & ~~~0.33~~~~~p & 1,2      & ~~~1.15 & DES~J045258.25--404011.5~~ & G  & 18.84   &   102.6 & 25.54 &   8 & FRIIncor             \\
J0457--4137~~~~~R & 074.2901 & --41.6327 & ~~2.76  & ~~~0.596~~~p  & 1        & ~~~1.10 & DES~J045709.61--413757.6~~ & G  & 21.07   &    21.3 & 25.45 &  35 & FRIIncor             \\
J0457--4445~~~~~R & 074.4563 & --44.7635 & ~~3.55  & ~~~0.581~~~p  & 1,2      & ~~~1.40 & DES~J045749.50--444548.4~~ & G  & 20.69   &    24.1 & 25.48 & 125 & FRIIasym             \\
J0458--4659~~~~~S & 074.5083 & --46.9964 & ~~3.25  & ~~~0.45~~~~~p & 1,2      & ~~~1.12 & DES~J045801.98--465947.1~~ & G  & 19.95   &   626.3 & 26.64 & 118 & FRIIrelic            \\
J0459--4637~~~~~R & 074.8534 & --46.6192 & ~~6.43  & ~~~0.310~~~p  & 1,2      & ~~~1.76 & DES~J045924.80--463709.1~~ & G  & 18.20   &    30.0 & 24.95 & 171 & FRIIrelic            \\
J0459--4507~~~~~S & 074.9130 & --45.1326 & ~~2.45  & ~~~0.62~~~~~p & 1        & ~~~1.00 & DES~J045939.12--450757.3~~ & G  & 20.75   &    74.7 & 26.04 &  23 & FRII                 \\
J0500--4242~~~~~R & 075.0310 & --42.7108 & ~~5.88  & ~~~1.1~~~~~~~p & 1,8,9   & ~~~2.88 & DES~J050007.44--424238.7~~ & Qc & 18.87   &    25.0 & 26.15 &  73 & FRIInaked            \\
J0502--4723C~~R   & 075.6359 & --47.3996 & ~~2.75  & ~~~0.94~~~~~p & 1        & ~~~1.30 & DES~J050232.61--472358.4~? & G  & 23.44   &    12.5 & 25.69 &  72 & FRII                 \\
J0507--4248~~~~~S & 076.9394 & --42.8065 & ~~3.7~  & ~~~0.46~~~~~p & 1,8      & ~~~1.29 & DES~J050745.45--424823.3~~ & Qc & 18.76   &    73.1 & 25.73 &  81 & FRII                 \\
J0508--4737C~~R   & 077.0647 & --47.6263 & ~~6.54  & ~~~0.421~~~p  & 1,2      & ~~~2.17 & DES~J050815.53--473734.5~~ & G  & 18.94   &    60.5 & 25.55 &  35 & FRII                 \\
J0509--4619~~~~~R & 077.3847 & --46.3188 & ~~2.74  & ~~~0.66~~~~~p & 1        & ~~~1.15 & DES~J050932.33--461907.6~~ & G  & 21.22   &    39.1 & 25.82 & 107 & FRII                 \\
J0519--4654~~~~~S & 079.7541 & --46.9148 & ~~3.8~  & ~~~0.5~~~~?~p & 1        & ~~~1.39 & DES~J051900.97--465453.3~? & G  & 21.48   &   114.2 & 26.00 & 123 & FRII                 \\
J0524--4235C~~R   & 081.0285 & --42.5939 & ~~6.16  & $>$0.65~~~?~p  & 1,8     & $>$2.56 & DES~J052406.82--423537.9~~ & Qc & 19.89   &    42.1 & 25.84 & 166 & FRII?                \\
J0524--4903~~~~~R & 081.1285 & --49.0535 & ~~2.28  & ~~~1.14~~~~~p & 1        & ~~~1.12 & DES~J052430.84--490312.5~~ & Qc & 23.88   &    25.8 & 26.20 &  42 & FRII?                \\
\bottomrule
\end{tabular}   \end{specialtable}
\begin{specialtable}[H]\ContinuedFloat
\widetable\small
\caption{{\em Cont.}}
\begin{tabular}{lcclllclllrcrl}

 ine
    ~~~~(1)          &    (2)       &   (3)         &   ~(4)    &    ~(5)  &   (6)  &     (7)  &    ~~~~~~~~~(8)    &   (9)   & (10)  &   (11)  &   (12)   &   (13)    &  (14)  \\
  Name, origin   &   RA$_J$     & Dec$_J$     &   LAS    & ~z, ztype & Ref(z) &  LLS     & ~~~~~~Hostname  & Host    & rmag   & S$_{888}$~ & log\,P$_{888}$ & RPA  & Radio\,Morphology \\
                & ($^{\circ}$) &  ($^{\circ}$) &  ~($'$)   &         &        &  (Mpc)   &           &  Type    & (mag)  &  (mJy)  &    (W/Hz)     &   ($^{\circ}$) &        \\
\midrule
J0528--4057~~~~~R & 082.1046 & --40.9601 & ~~2.4~  & ~~~0.9~~~~?~p  & 1,8     & ~~~1.12 & DES~J052825.10--405736.3~~ & Qc & 23.24   &    32.9 & 26.07 & 167 & FRII                 \\
J0532--4118~~~~~R & 083.2079 & --41.3158 & ~~2.74  & ~~~0.72~~~~~p & 1        & ~~~1.19 & DES~J053249.90--411856.8~~ & G  & 20.94   &    22.1 & 25.66 &  44 & FRII,WAT?            \\
J0533--4907~~~~~S & 083.4639 & --49.1298 & ~~3.25  & ~~~0.50~~~~~p & 1,2      & ~~~1.19 & DES~J053351.34--490747.3~~ & G  & 19.93   &    85.7 & 25.88 &  38 & FRIIasym             \\
J0533--4312~~~~~R & 083.4847 & --43.2054 & ~~4.15  & ~~~0.70~~~~~p & 1        & ~~~1.78 & DES~J053356.31--431219.5~~ & G  & 21.33   &    36.4 & 25.85 & 102 & FRIIncor             \\
J0545--4350~~~~~R & 086.4777 & --43.8415 & ~~4.88  & ~~~0.64~~~~~p & 1        & ~~~2.01 & DES~J054554.64--435029.2~~ & G  & 20.47   &    41.8 & 25.82 &  59 & FRIIncor             \\
J0550--4014~~~~~R & 087.6864 & --40.2486 & ~~3.1~  & ~~~0.9~~~~?~p & 1,8      & ~~~1.45 & DES~J055044.73--401455.1~~ & Qc & 21.65   &    48.0 & 26.23 &  40 & FRIIrelic            \\
J0551--4519~~~~~R & 087.7760 & --45.3332 & ~~2.3~  & ~~~1.07~~~~~p & 1        & ~~~1.12 & DES~J055106.24--451959.6~~ & G  & 22.59   &    59.3 & 26.50 &  81 & FRII                 \\
J0551--4902~~~~~R & 087.9728 & --49.0364 & ~~3.0~  & ~~~1.31~~~~~p & 1        & ~~~1.51 & DES~J055153.46--490211.0~~ & G  & 24.25   &    10.3 & 25.95 &  50 & FRII                 \\
J0554--4103~~~~~S & 088.6400 & --41.0624 & ~~3.00  & ~~~0.87~~~~~p & 1        & ~~~1.39 & DES~J055433.59--410344.5~~ & G  & 22.34   &   162.5 & 26.73 & 109 & FRII                 \\
J0600--4908~~~~~R & 090.1793 & --49.1469 & ~~3.55  & ~~~1.2~~~~~~~p & 1       & ~~~1.77 & DESI~J090.1793--49.1469 ~? & G? & 23.49   &    61.2 & 26.63 &  79 & FRIIasym ?           \\
J0604--4459~~~~~R & 091.1750 & --44.9970 & ~~3.07  & ~~~0.78~~~~~p & 1        & ~~~1.37 & DES~J060441.98--445949.1~~ & G  & 22.29   &    72.6 & 26.26 & 158 & FRII                 \\
J0609--4518~~~~~R & 092.3167 & --45.3051 & ~~5.99  & ~~~0.513~~~p  & 1        & ~~~2.22 & DES~J060915.99--451818.2~~ & G  & 19.37   &    52.0 & 25.69 &   2 & FRII                 \\
J0618--4628~~~~~S & 094.7272 & --46.4721 & ~~7.32  & ~~~0.39~~~~~p & 1,2      & ~~~2.32 & DES~J061854.52--462819.3~~ & G  & 19.57   &   252.3 & 26.10 &   9 & FRII                 \\
J2022--4156~~~~~R & 305.5250 & --41.9452 & ~~2.52  & ~~~0.800~~~p  & 1        & ~~~1.14 & DES~J202206.00--415642.8~~ & Qc & 21.31   &    40.5 & 26.03 &  74 & FRII                 \\
J2022--4618~~~~~R & 305.5850 & --46.3011 & ~~3.6~  & ~~~0.35~~?~p  & 1,2      & ~~~1.06 & DES~J202220.40--461803.8~? & Qc & 20.00   &    33.7 & 25.12 &  14 & FRIIncor,asym        \\
J2024--4947~~~~~R & 306.2337 & --49.8000 & ~~2.65  & ~~~0.86~~~~~p & 1        & ~~~1.22 & DES~J202456.08--494759.8~? & G  & 21.77   &    17.9 & 25.75 & 174 & FRIIrelic,ncor       \\
J2025--4003~~~~~R & 306.2775 & --40.0619 & ~~6.0~  & ~~~0.57~~~~~p & 1        & ~~~2.35 & DES~J202506.58--400342.8~~ & G  & 21.08   &   185.2 & 26.35 &  10 & FRII                 \\
J2029--4119~~~~~S & 307.4041 & --41.3174 & ~~4.0~  & ~~~0.355~~~p  & 1,2      & ~~~1.20 & DES~J202936.98--411902.7~~ & G  & 18.45   & 1350.~~~& 26.73 & 148 & FRII                 \\
J2041--4546~~~~~R & 310.3727 & --45.7767 & ~~2.8~  & ~~~0.82~~~~~p & 1        & ~~~1.27 & DES~J204129.45--454636.2~~ & G  & 20.91   &   100.2 & 26.45 &  22 & FRII                 \\
J2042--4801~~~~~R & 310.5939 & --48.0302 & ~~2.2~  & ~~~0.83~~~~~p & 1        & ~~~1.00 & DES~J204222.52--480148.8~~ & G  & 21.75   &    43.2 & 26.10 &  60 & FRIIrelic            \\
J2045--4340~~~~~R & 311.2527 & --43.6793 & ~~3.5~  & ~~~0.978~~~p  & 1        & ~~~1.67 & DES~J204500.64--434045.5~~ & G  & 23.33   &   135.3 & 26.77 &  66 & FRIIrelic            \\
J2049--4014~~~~~R & 312.2829 & --40.2416 & ~~3.8~  & ~~~0.475~~~p  & 1,2      & ~~~1.35 & DES~J204907.89--401429.7~~ & G  & 20.01   &    10.9 & 24.93 & 125 & FRIIrelic            \\
J2050--4647~~~~~S & 312.5283 & --46.7885 & ~~4.91  & ~~~0.55~~~~~p & 1,2      & ~~~1.89 & DES~J205006.79--464718.5~~ & G  & 20.76   &    74.9 & 25.92 &   0 & FRII                 \\
J2052--4813~~~~~E & 313.0764 & --48.2225 & ~~4.91  & ~~~0.395~~~p  & 1,2      & ~~~1.57 & DES~J205218.33--481321.2~~ & G? & 19.71   &    55.4 & 25.45 &  77 & FRII                 \\
J2054--4431~~~~~R & 313.7169 & --44.5245 & ~~3.0~  & ~~~0.55~~~~~p & 1        & ~~~1.16 & DES~J205452.05--443128.3~~ & G  & 20.50   &    12.1 & 25.13 &  34 & FRIIrelic            \\
J2055--4413C~~R   & 313.8971 & --44.2268 & ~~3.1~  & ~~~0.399~~~p  & 1,2      & ~~~1.00 & DES~J205535.29--441336.5~~ & G  & 19.25   &     6.3 & 24.52 & 179 & FRIIrelic            \\
J2056--4821~~~~~E & 314.0627 & --48.3567 & ~~7.0~  & ~~~0.35~~~~~p & 1,2,8    & ~~~2.07 & DES~J205615.04--482123.9~~ & G  & 19.14   &    10.2 & 24.60 &  98 & FRII                 \\
J2056--4845~~~~~E & 314.2441 & --48.7567 & ~~2.35  & ~~~0.94~~~~~p & 1        & ~~~1.11 & DES~J205658.59--484524.4~~ & G  & 22.43   &    21.3 & 25.92 &  21 & FRII                 \\
J2108--4336~~~~~R & 317.1310 & --43.6134 & ~~2.95  & ~~~1.06~~~~~p & 1        & ~~~1.44 & DES~J210831.43--433648.3~~ & G  & 22.06   &   124.6 & 26.81 &  84 & FRII                 \\
J2110--4112~~~~~R & 317.5495 & --41.2031 & ~~3.9~  & ~~~0.392~~~p  & 1,2      & ~~~1.24 & DES~J211011.87--411211.1~~ & G  & 18.50   &    56.1 & 25.45 &   0 & FRIIbent,WAT?        \\
J2111--4948~~~~~E & 317.9387 & --49.8122 & ~~4.49  & ~~~0.283~~~p  & 1,2      & ~~~1.15 & DES~J211145.28--494843.9~~ & G  & 18.70   &    24.3 & 24.77 &  32 & FRII                 \\
J2117--4415~~~~~R & 319.3348 & --44.2578 & ~~3.2~  & ~~~0.7~~~~?~p & 1,2,8,9  & ~~~1.37 & DES~J211720.35--441528.1~~ & Qc & 19.34   &    66.3 & 26.11 &  23 & FRII                 \\
J2138--4435C~~R   & 324.6969 & --44.5844 & ~~4.39  & ~~~1.06~~~~~p & 1        & ~~~2.14 & DES~J213847.26--443503.7~~ & G  & 22.52   &    13.0 & 25.83 & 114 & FRII                 \\
J2144--4818~~~~~E & 326.0088 & --48.3159 & ~~7.95  & ~~~0.136~~~p  & 1,2      & ~~~1.15 & 2MASX~J21440210--4818581~~ & G  & 17.03   &   191.2 & 24.96 & 167 & FRIIremn,ncor        \\
J2149--4226~~~~~R & 327.3891 & --42.4441 & ~~2.63  & ~~~0.66~~~~~p & 1,2,8,9  & ~~~1.12 & DES~J214933.38--422638.7~~ & GQ & 19.56   &   402.0 & 26.83 & 163 & FRII                 \\
J2152--4911~~~~~E & 328.1957 & --49.1952 & ~~2.95  & ~~~0.58~~~~~p & 1,2      & ~~~1.16 & DES~J215246.97--491142.9~? & G  & 20.05   &    30.7 & 25.58 & 102 & FRII                 \\
J2154--4552~~~~~S & 328.5959 & --45.8753 & ~~7.2~  & ~~~0.1453~s   & 7        & ~~~1.10 & 2MASX~J21542297--4552315~~ & G  & 16.12   & 1106.~~~& 25.79 &  40 & FRIIasym             \\
\bottomrule
\end{tabular}   \end{specialtable}
\begin{specialtable}[H]\ContinuedFloat
\widetable\small
\caption{{\em Cont.}}
\begin{tabular}{lcclllclllrcrl}

 ine
    ~~~~(1)          &    (2)       &   (3)         &   ~(4)    &    ~(5)  &   (6)  &     (7)  &    ~~~~~~~~~(8)    &   (9)   & (10)  &   (11)  &   (12)   &   (13)    &  (14)  \\
  Name, origin   &   RA$_J$     & Dec$_J$     &   LAS    & ~z, ztype & Ref(z) &  LLS     & ~~~~~~Hostname  & Host    & rmag   & S$_{888}$~ & log\,P$_{888}$ & RPA  & Radio\,Morphology \\
                & ($^{\circ}$) &  ($^{\circ}$) &  ~($'$)   &         &        &  (Mpc)   &           &  Type    & (mag)  &  (mJy)  &    (W/Hz)     &   ($^{\circ}$) &        \\
\midrule
J2155--4235~~~~~S & 328.8819 & --42.5867 & ~~3.47  & $>$0.70~~~~~p & 8,9,14   & $>$1.49 & DES~J215531.65--423512.0~~ & Qc & 18.46   &   132.5 & 26.41 &  38 & FRII                 \\
J2202--4004~~~~~R & 330.6272 & --40.0686 & ~~5.1~  & ~~~0.86~~~~~p & 1        & ~~~2.35 & DES~J220230.53--400406.8~? & G  & 21.62   &    39.1 & 26.09 & 166 & FRII                 \\
J2205--4810~~~~~R & 331.4580 & --48.1686 & ~~2.65  & ~~~0.60~~~~~p & 1        & ~~~1.06 & DES~J220549.91--481007.0~~ & GP & 21.47   &    16.3 & 25.34 & 161 & FRIIrelic            \\
J2215--4225~~~~~R & 333.8671 & --42.4300 & ~~3.5~  & ~~~0.42~~~~~p & 1        & ~~~1.16 & DES~J221528.11--422548.0~~ & G  & 18.75   &     8.1 & 24.68 &  22 & FRIIremn             \\
J2216--4328~~~~~R & 334.2185 & --43.4674 & ~~2.05  & ~~~1.6~~~~?~p & 1,8      & ~~~1.04 & DES~J221652.36--432802.2~? & Qc & 24.60   &     5.0 & 25.84 & 164 & FRII                 \\
J2218--4959~~~~~E & 334.5944 & --49.9908 & ~~2.4~  & ~~~0.8~~~~?~p & 1        & ~~~1.03 & DES~J221822.66--495927.0~~ & G  & 24.35   &    10.0 & 25.43 &  78 & FRII                 \\
J2219--4748~~~~~R & 334.7847 & --47.8047 & ~~3.1~  & ~~~0.46~~~~~p & 1,2      & ~~~1.08 & DES~J221908.32--474816.8~~ & Qc & 18.78   &    10.3 & 24.88 & 172 & FRIIrelic            \\
J2220--4854~~~~~E & 335.1798 & --48.9048 & ~~2.53  & ~~~0.80~~~~~p & 1        & ~~~1.14 & DES~J222043.14--485417.1~~ & G  & 20.42   &    16.1 & 25.64 & 125 & FRII                 \\
J2224--4724~~~~~S & 336.0727 & --47.4009 & ~~3.7~  & ~~~0.85~~~~~p & 1        & ~~~1.70 & DES~J222417.45--472403.2~~ & G  & 22.66   &    96.7 & 26.48 &  85 & FRII                 \\
J2226--4316~~~~~R & 336.6399 & --43.2765 & ~13.8~  & ~~~0.09309\,s & 12       & ~~~1.43 & 2MASX~J22263358--4316356~~ & G  & 17.75   &   170.~~& 24.56 & 153 & FRIIremn             \\
J2226--4624C~~R   & 336.7430 & --46.4036 & ~~2.62  & ~~~1.40~~~~~p & 1        & ~~~1.33 & DES~J222658.31--462413.1~~ & Qc & 23.88   &    44.9 & 26.66 &  11 & FRIIncor             \\
J2228--4210~~~~~S & 337.0102 & --42.1740 & ~~4.94  & ~~~0.55~~?~p  & 1,2      & ~~~1.90 & DES~J222802.44--421026.4~~ & G  & 20.51   &    93.8 & 26.02 & 171 & FRII                 \\
J2232--4025~~~~~R & 338.1574 & --40.4270 & ~~3.78  & ~~~0.68~~~~~p & 1        & ~~~1.60 & DES~J223237.77--402537.2~~ & G  & 21.21   &    12.8 & 25.37 &  25 & FRIIrelic            \\
J2236--4113~~~~~R & 339.2012 & --41.2225 & ~~4.56  & ~~~0.3~~~~~~~p & 1,2,9   & ~~~1.22 & DES~J223648.28--411321.0~~ & G  & 19.12   &    61.0 & 25.22 &  13 & FRII                 \\
J2239--4450~~~~~S & 339.8511 & --44.8384 & ~~3.4~  & ~~~0.42~~~~~p & 1        & ~~~1.13 & DES~J223924.27--445018.1~~ & G  & 19.84   &    34.8 & 25.31 & 156 & FRII                 \\
J2240--4724~~~~~S & 340.1112 & --47.4028 & ~12.~~  & ~~~0.09161\,s & 7        & ~~~1.23 & 2MASX~J22402674--4724100~~ & G  & 15.41   &    53.~~& 24.04 &  25 & FRIIremn,DDRG?       \\
J2245--4239~~~~~R & 341.4210 & --42.6533 & ~~2.54  & ~~~1.00~~~~~p & 1        & ~~~1.22 & DES~J224541.03--423911.8~~ & G  & 21.48   &    15.5 & 25.85 &   7 & FRII                 \\
J2246--4934~~~~~R & 341.7351 & --49.5830 & ~~3.4~  & ~~~0.80~~~~~p & 1        & ~~~1.53 & DES~J224656.42--493458.9~~ & G  & 21.87   &    27.7 & 25.87 & 170 & FRII                 \\
J2247--4034~~~~~A & 341.7816 & --40.5705 & ~~2.75  & ~~~0.53~~~~~p & 1,8,9,14 & ~~~1.04 & DES~J224707.59--403413.7~~ & Qc & 19.72   &    64.5 & 25.82 &  16 & FRII                 \\
J2247--4753~~~~~R & 341.9390 & --47.8977 & ~~3.6~  & ~~~0.34~~~~~p & 1        & ~~~1.05 & DES~J224745.37--475351.8~~ & G  & 21.69   &     7.0 & 24.41 &  30 & FRII                 \\
J2250--4751~~~~~S & 342.6531 & --47.8643 & ~~4.04  & ~~~0.32~~~~~p & 1,2      & ~~~1.13 & DES~J225036.75--475151.5~~ & G  & 18.84   &    22.6 & 24.85 & 121 & FRII                 \\
J2253--4424~~~~~R & 343.2764 & --44.4038 & ~~2.53  & ~~~1.07~~~~~p & 1        & ~~~1.23 & DES~J225306.34--442413.6~? & G  & 22.25   &    63.1 & 26.53 & 175 & FRIIncor             \\
J2258--4819~~~~~R & 344.5067 & --48.3254 & ~~3.0~  & ~~~0.98~~~~~p & 1        & ~~~1.43 & DES~J225801.59--481931.2~~ & G  & 22.99   &    17.3 & 25.87 &  63 & FRII                 \\
J2301--4359~~~~~R & 345.4059 & --43.9971 & ~~8.0~  & ~~~0.11685\,s & 7        & ~~~1.01 & 2MASX~J23013741--4359497~~ & G  & 15.33   &   160.6 & 24.75 &   7 & FRI/IIrelic,precess  \\
J2306--4045~~~~~S & 346.6868 & --40.7622 & ~~5.24  & ~~~0.28~~~~~p & 1,2      & ~~~1.33 & DES~J230644.84--404543.9~~ & G  & 19.31   &    46.9 & 25.04 &  94 & FRIIremn,ncor        \\
J2307--4049~~~~~R & 346.8101 & --40.8251 & ~~3.3~  & ~~~0.71~~~~~p & 1        & ~~~1.42 & DES~J230714.42--404930.5~~ & G  & 21.66   &    16.0 & 25.51 & 179 & FRII                 \\
J2317--4645~~~~~R & 349.3573 & --46.7614 & ~~2.7~  & ~~~0.80~~~~~p & 1        & ~~~1.22 & DES~J231725.76--464540.8~~ & Qc & 20.40   &    23.2 & 25.79 &  55 & FRII                 \\
J2319--4449~~~~~R & 349.8362 & --44.8181 & ~~2.8~  & ~~~0.76~~~~~p & 1        & ~~~1.24 & DES~J231920.68--444905.1~~ & G  & 20.94   &    33.5 & 25.90 &  22 & FRII                 \\
J2325--4311~~~~~R & 351.2995 & --43.1922 & ~~4.48  & ~~~1.03~~~~~p & 1        & ~~~2.17 & DES~J232511.86--431131.9~~ & G  & 22.82   &    42.2 & 26.31 &  85 & FRII                 \\
J2325--4153~~~~~R & 351.3907 & --41.8939 & ~~3.0~  & ~~~0.51~~~~~p & 1        & ~~~1.11 & DES~J232533.76--415338.0~~ & G  & 20.22   &    21.6 & 25.30 & 122 & FRIIremn,ncor        \\
J2331--4928~~~~~R & 352.9503 & --49.4797 & ~~2.44  & ~~~1.24~~~~~p & 1        & ~~~1.22 & DESI~J352.9504--49.4797 ~~ & G  & 24.05   &    66.2 & 26.70 &  49 & FRIIrelic            \\
J2345--4754~~~~~R & 356.3691 & --47.9028 & ~~2.5~  & ~~~0.74~~~~~p & 1        & ~~~1.09 & DES~J234528.57--475410.1~~ & G  & 21.27   &    15.1 & 25.53 & 165 & FRI/II               \\
J2345--4642~~~~~R & 356.4543 & --46.7158 & ~~4.2~  & ~~~0.56~~~~~p & 1        & ~~~1.63 & DES~J234549.04--464256.8~~ & G  & 19.98   &    76.6 & 25.95 &  67 & FRIIncor             \\
J2347--4047~~~~~R & 356.9115 & --40.7904 & ~~4.0~  & ~~~0.50~~~~~p & 1        & ~~~1.46 & DES~J234738.75--404725.5~~ & Qc & 22.61   &    12.7 & 25.05 & 156 & FRII                 \\
J2347--4837~~~~~R & 356.9846 & --48.6326 & ~~2.48  & ~~~1.35~~~~~p & 1,8,9    & ~~~1.25 & DES~J234756.29--483757.3~~ & Qc & 18.13   &    21.3 & 26.30 & 139 & FRIIrelic            \\
J2350--4232~~~~~S & 357.5374 & --42.5375 & ~~2.6~  & ~~~0.99~~~~~p & 1        & ~~~1.25 & DES~J235008.98--423215.1~~ & G  & 22.59   &    25.1 & 26.05 & 167 & FRIIrelic            \\
J2357--4010~~~~~R & 359.3201 & --40.1831 & ~~3.55  & ~~~0.50~~~~~p & 1        & ~~~1.29 & DES~J235716.83--401059.0~~ & G  & 20.34   &    48.7 & 25.64 &  81 & FRIIrelic            \\
\bottomrule
\end{tabular}   
\end{specialtable}
\finishlandscape


\noindent
in Table~\ref{tab:morph}), and one (J0023--4732) shares both of the latter properties.
Apart from these clear FR\,IIs we found 11 (6.2\%) mixed types I/II, and
only a single one (J0422--4518C) is of type FR\,I and classified as a
candidate WAT with very bent, diffuse, and low surface brightness lobes.
We tagged 21 of the GRS as ``remnant" and 34 as ``relic" sources which
refers to the degree of diffuseness of the lobes. These will be discussed
in Sect.~\ref{sect:rrgs}. Since we use the tag ``relic" only in connection with
the tag FR\,II, there should not be any confusion with ``relic" radio
sources found in clusters of galaxies and not associated with individual
host galaxies.
Nine FR\,IIs were labelled as ``asymmetric" implying that the most likely
host was found more than $\sim$2 times closer to one of the outer lobes
than the other. This is not unusual given that the ratio of the larger
to smaller distance of hotspots of FR\,IIs to their host can reach and
exceed five \cite{delarosa19}.
Apart from the known double-double GRS (J0116--4722) only one other noteworthy
example (J2240--4724) of this type of source was found.
No clear examples of hybrid morphology sources (see e.g.~\cite{kapinska17})
were found except for possibly J0138--4116 and J0433--4948, and no 
single example of the so-called ``Odd radio circles" (ORC, \cite{norris21a})
was found in our inspection of RACS.

While the original FR classification \cite{fr74} suggested a
clear-cut separation of these types in radio luminosity, it was later shown
that the dividing radio luminosity not only increases with optical luminosity
\cite{owen91, owen94} but also that there is a large overlap in radio luminosities
between these types (e.g.\ \cite{best09, mingo19}).  In our new sample of GRS
we can see only a marginal trend in the median $log\,P_{888}$ from 25.0 for
the 12 FR\,I or FR\,I/II sources, 25.7 for the 51 FRIIrelic or remnant sources,
and 26.0 for the 76 clear FR\,II sources.  We defer any discussion of this
so-called ``Radio-HR" \cite{owen93} or Owen--Ledlow diagram \cite{owen94}
to a later paper, including smaller sources with a more appreciable fraction of
FR\,I types.

As noted in Sect.~\ref{sect:fluxint} our algorithm for flux integration
includes an estimate of equipartition parameters, assuming $k=1$ 
for the proton-to-electron energy ratio, a filling factor of 1,
a uniform source magnetic field in the plane of the sky, and 
$\alpha=-0.8$ for the source's spectral index \cite{miley80}. 
For the source lobes, excluding the core emission, we find magnetic 
fields, $B_{eq}$, of $\sim$0.9--3.0\,$\mu$G (0.09--0.3\,nT) independent of
using RACS or CRACS, and minimum energy densities in particles and fields,
$u_{min}$, of $\sim$1.4--17$\times10^{-14}$J\,m$^{-3}$ for RACS, and
0.8--14$\times10^{-14}$J\,m$^{-3}$ for CRACS. The minimum total energies
range from 1.4$\times10^{50}$\,J to 1.7$\times10^{54}$\,J for RACS and
from 2$\times10^{50}$\,J to 2.2$\times10^{54}$\,J for CRACS.
The median values are comparable to those found for 3C\,236 by \cite{willis74}.
We defer a more detailed discussion of this to a later paper.

In what follows we select a few examples of the variety of GRS we found
by means of overlays of CRACS contours on $g,r,i$-band composites
from DES\,DR1. We recommend viewing these images at an amplified scale
to appreciate the optical host and its environment in more detail.

\subsubsection{Radio Sources exceeding an LLS of 2\,Mpc} \label{sect:2mpc}

\begin{figure}[H] 
\includegraphics[width=6.3cm]{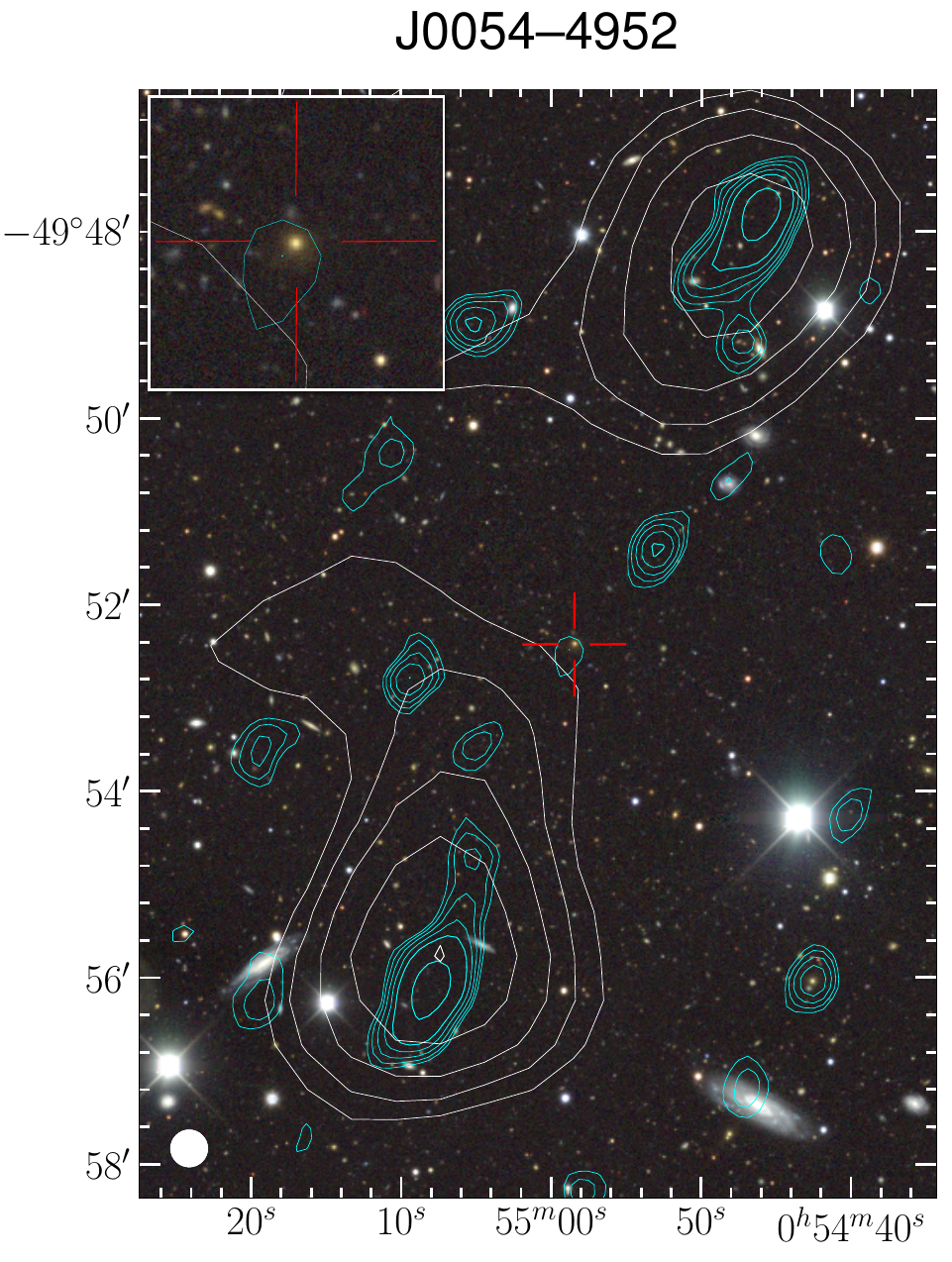} 
\includegraphics[width=7.06cm]{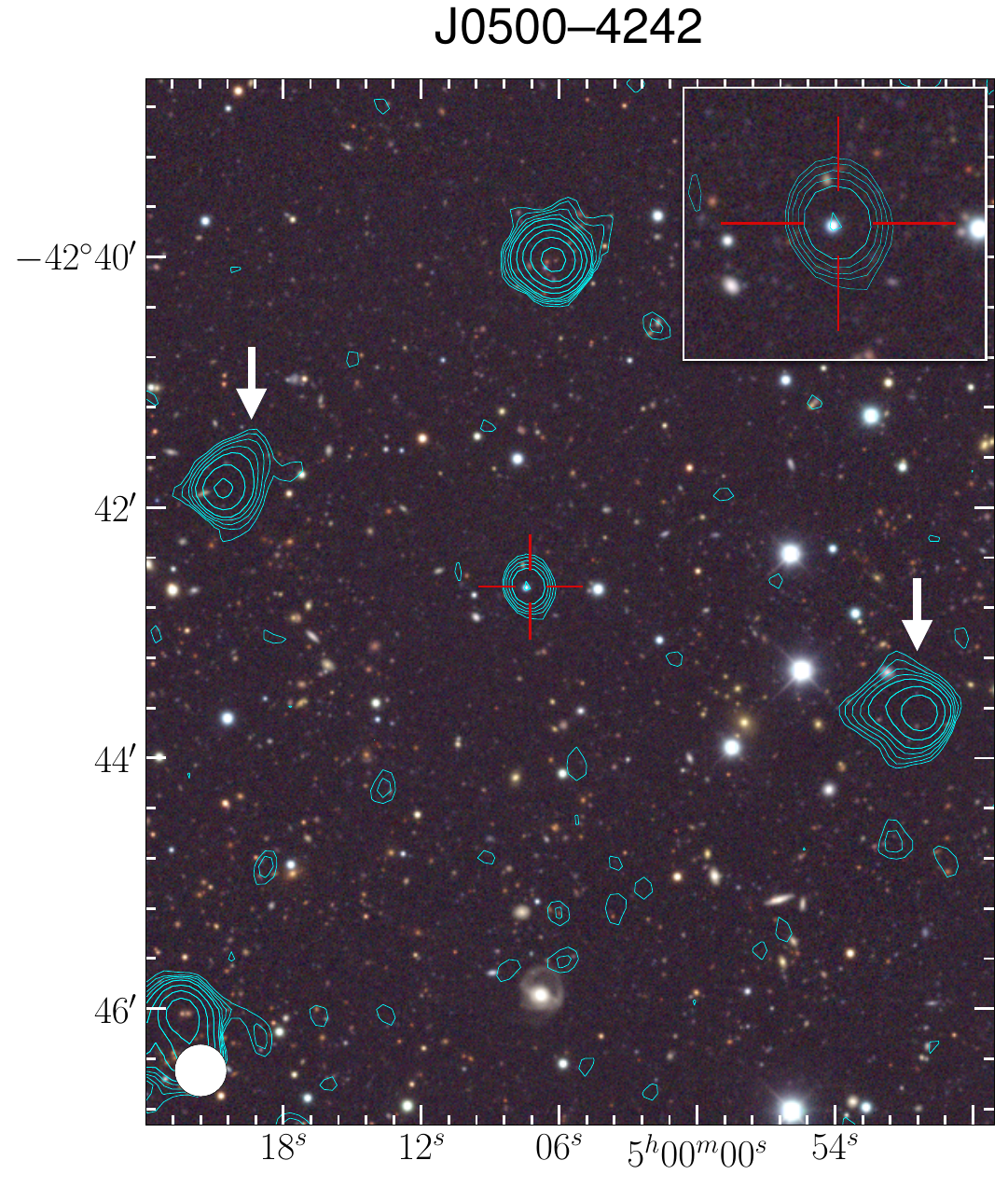} 
\caption{The two largest GRS in our sample: the GRG J0054--4952
of 3.4\,Mpc (left), and the GRQ J0500--4242 of 2.9\,Mpc (right), overlaid on a
$g,r,i$-composite of optical images from DES\,DR1. Green contours from CRACS
are plotted at  $\rm 2 \times [1, 2^{1/2}, 2^{1}, 2^{3/2}, 2^2, ...]\times rms$.
The gapped red crosses at centre mark the location of the hosts. 
The insets show a zoom of an area of 1$'\times1'$ around the hosts.
The left panel shows additional contours from GLEAM \cite{hurleywalker17}
in the 170--231\,MHz band, plotted at the same contour intervals, 
but with rms=6\,mJy\,beam$^{-1}$ for a beam size of $\sim2'$. These delineate the
outer lobes, while in the right panel the lobes are indicated with white
arrows.}
\label{fig:2mpc}
\end{figure}

The previously published compilations of 458~GRS (see Sect.~\ref{sect:compar})
comprise six GRS (1.3\%) with LLS$>$3\,Mpc and 52 (11.4\%) larger than 2\,Mpc.
We find similar fractions in our sample of 178 new GRS: there is
one GRS (0.6\%) larger than 3\,Mpc and 18 (10\%) larger than 2\,Mpc.
Our largest GRS is J0054--4952, hosted by the $r$=19.4\,mag galaxy
DES~J005458.45--495226.0 at $z_{phot}$=0.47 (based on values of 0.46 and 0.49 
from \cite{bilicki16} and \cite{zhou21}), giving an LLS of 3.4\,Mpc.
The second-largest is J0500--4242, hosted by the $r$=18.9\,mag QSO candidate
DES~J050007.44--424238.7 at $z_{phot}$=1.1 (based on values of 0.85, 1.1,
and 1.5 from \cite{zhou21,krogager18,flesch21}), giving an LLS of 2.9\,Mpc.
In the six literature compilations (Sect.~\ref{sect:compar}) there
is only one GRS with both $z>1$ and LLS$>$2\,Mpc, the QSO FBQS~J0204--0944 at
$z_{spec}$=1.004 with LLS=2.1\,Mpc, making our new find of J0500--4242
the largest GRQ known at $z>1$. We classify the latter as ``FR\,IInaked"
since the outer lobes are barely resolved with little or no indication of
a radio tail or bridge pointing towards the host. Future higher-resolution
and sensitivity radio observations should show some extended structure
around these hotspots.  Both these GRS are shown in Fig.~\ref{fig:2mpc}.

\subsubsection{Asymmetric Sources and Wide-Angle Tails (WATs)} \label{sect:wats}

\begin{figure}[H] 
\includegraphics[width=6.75cm]{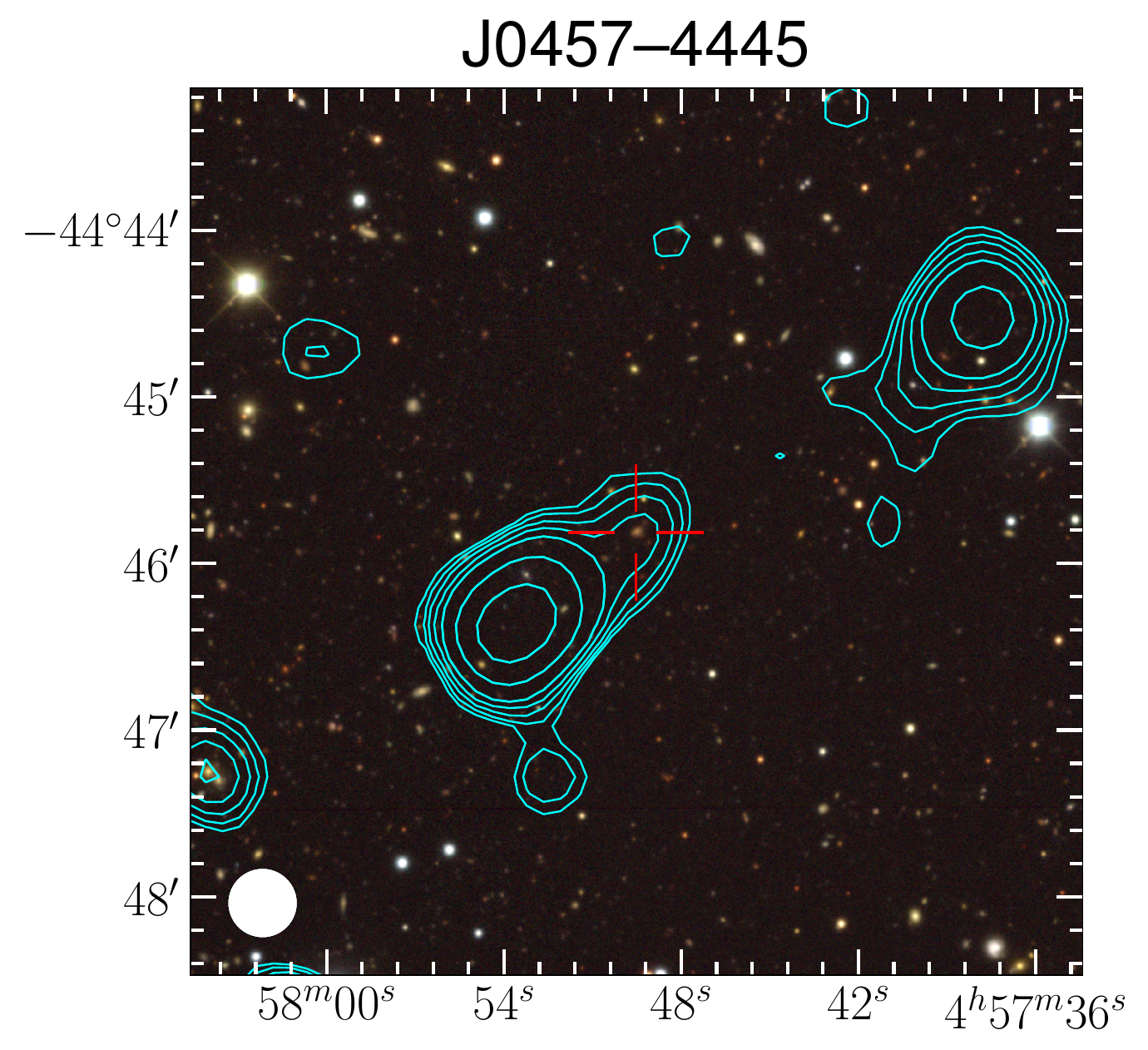}
\includegraphics[width=6.75cm]{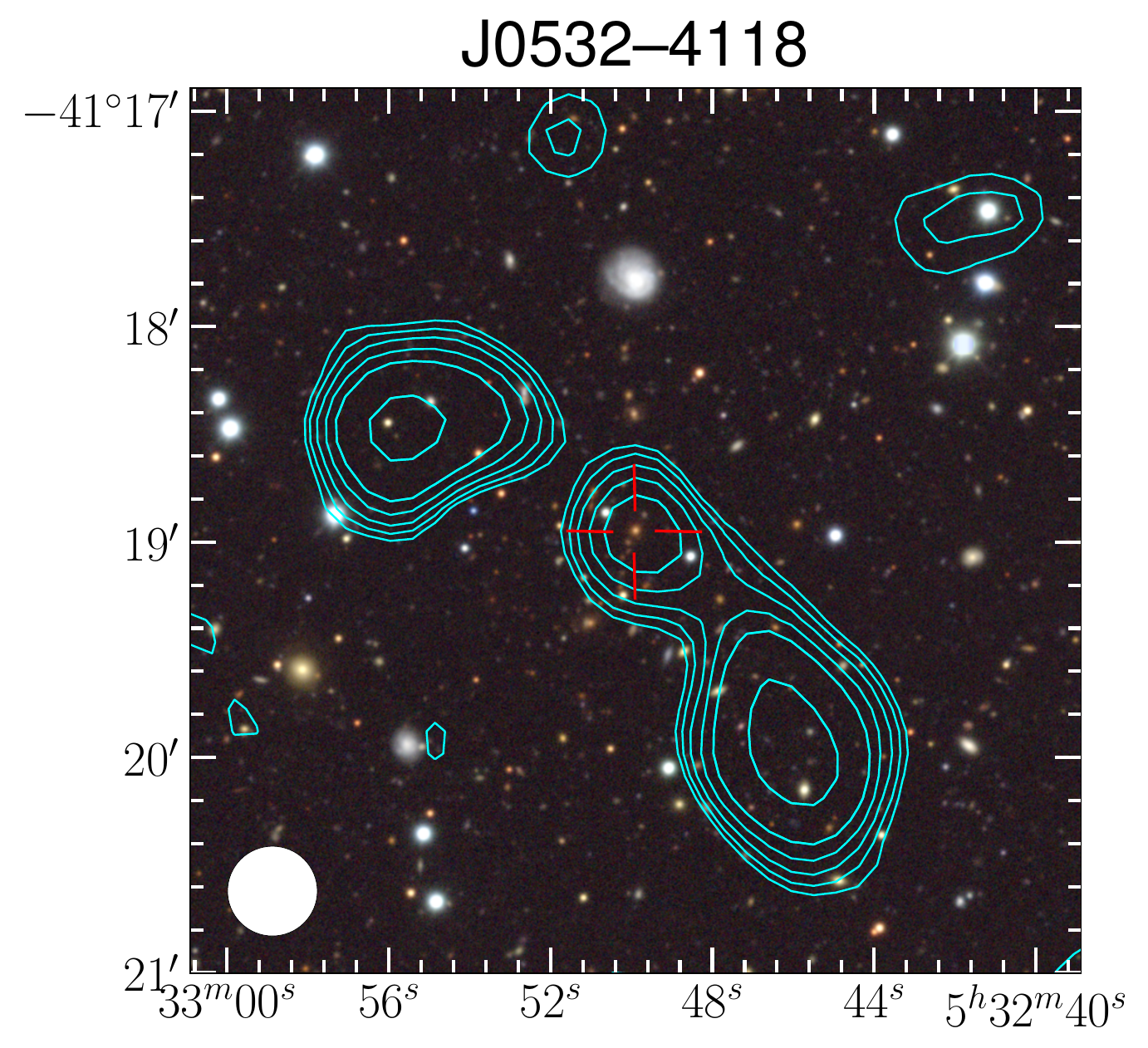} 
\caption{Left: an example of an asymmetric GRS J0457--4445 of
size 1.4\,Mpc; right: the WAT-type source J0532--4118 in a z$\sim0.7$ cluster.
Background image, contour levels,and gapped red cross are as in Fig.~\ref{fig:2mpc}.}
\label{fig:asym_wats}
\end{figure}

About nine of our newly found GRS have ratios of their opposite lobe lengths
of about two. The left panel of Fig.~\ref{fig:asym_wats} shows the example of
J0457--4445 hosted by the $r$=20.7\,mag galaxy  DES J045749.50--444548.4
at $z_{phot}$=0.58 giving an LLS of 1.4\,Mpc.

WATs tend to occur exclusively
in clusters of galaxies and are predominantly hosted by the brightest
cluster member. Although they are generally considered of FR\,I type,
they appear as either jet-dominated FR\,I types or lobe-dominated
FR\,II types or a mixture thereof.
There are no clear-cut examples of giant WATs in our sample, and
the object that comes closest to this morphology is
J0532--4118 at $z_{phot}$=0.72 with LLS$\sim$1.2\,Mpc, hosted by
the $r=20.9$\,mag galaxy DES~J053249.90--411856.8 (right panel of 
Fig.~\ref{fig:asym_wats}).
The DES image shows indications of a distant cluster, in fact, listed as
ID~4063500118 by \cite{zou21}, and as WaZP\,DES\,YR1~J053249.9--411856
in \cite{aguena21}, with the GRG host being the brightest member
in $i$-band and the third-brightest in $r$-band (the brightest in $r$
is located 33$''$ N of the WAT host). WAT type sources are not expected to
grow to sizes larger than about a Mpc, since they tend to be located
near the density peak of the intracluster medium of their host cluster.
Thus, J0532--4118 may be exceptional, also for its high redshift.
It is difficult to identify the largest WAT reported in literature
given that this morphological type is rarely mentioned
in publications on GRS, but e.g.\ J2233+1315 hosted by the $r=$15.2\,mag,
$z_{phot}$=0.093 galaxy 2MASX~J22330133+1315019 which is the brightest
of cluster WHL~J223301.3+131503 extends over 14.7$'$ or $\sim$1.7\,Mpc
\cite{dabhade17} \footnote{Note that this
galaxy does not have a $z_{spec}$ neither in \cite{sdssdr16} nor as 
claimed by NED which quotes
$z_{spec}$=0.093 from \cite{dabhade17} which is actually a $z_{phot}$.
Simbad (\url{http://simbad.u-strasbg.fr}) quotes $z_{spec}$=0.1021 from
\cite{yuan16}, obtained from \cite{wen15} who assigned this redshift to
the cluster, but the redshift refers to another cluster member, and
the radial velocity of 34851\,km\,s$^{-1}$ in \cite{condon19} cannot
be traced to any literature reference.},
or else the FR\,I source J1049+5513 listed in \cite{capetti17}
and hosted by 2MASX~J10490732+5513153 at $z_{spec}$=0.1262 can be
seen to extend over 11.5$'$ or 1.56\,Mpc in the Lockman Hole
LoTSS Deep Field \cite{tasse21}. Since the radio morphology of the latter
is quite likely foreshortened by projection, surveys like LoTSS or
EMU should reveal even larger WATs that are oriented closer
to the plane of the sky.

\subsubsection{Giant Radio Sources associated with Clusters of Galaxies}   \label{sect:clust}

Although the visual classification of the environment of the GRS hosts
in our sample is beyond the scope of this work, we cross-matched the
181~GRS of Table~\ref{tab:grglist} with major cluster catalogues and found
the following coincidences.

The host galaxy of J0406--4544 is the brightest and dominant member of
cluster WHY~J040607.9-454451 at $z_{phot}$=0.315 \cite{wen18}. The radio
source is a symmetric and straight FR\,II of projected size 1.8\,Mpc,
with its radio axis oriented perpendicular to the optical major axis
of the outer halo of the host galaxy, but $\sim25^{\circ}$ off from
perpendicular for the innermost isophotes, suggesting the presence of
isophote twist.  The cluster is rather poor and scattered, and as yet
unreported as X-ray emitter.

A cross-match of our 181 GRG hosts with 2.52 million member galaxies of
60542 clusters of the WaZP cluster catalogue \cite{aguena21} yielded 14 matches,
of which two are unlikely cluster members, seven are brightest members
(one is the above-mentioned J0532--4118), and another five are
lower-ranked members.  The six other brightest cluster members
(J0020--4625, J0105--4505, J0213--4744, J0429--4517, J0508--4737, and J2250--4751)
are all hosts of fairly straight FR\,II type GRS, which is considered rare.
All five lower-ranked cluster member GRS hosts are also FR\,IIs, and
special mention deserves the GRS J0150--4507, which is the 2nd-brightest
member of the very rich and filamentary cluster WaZP\,DES\,YR1~J015035.3--451112
at $z_{phot}\sim$0.3,
also detected via the Syunyaev-Zeldovich (SZ) effect as PSZ2~G273.69--68.38
and SPT-CL~J0150--4511.
The GRS host lies in a filament $\sim7'$ ($\sim$1.9\,Mpc) NW of the cluster 
centre, with its radio source axis oriented perpendicular to the galaxy 
filament, similar to the trend seen by \cite{malarecki13, malarecki15}.

In a cross-match of our 181 GRG hosts with 12.0 million member galaxies of
540,432 clusters from the DESI Legacy Imaging Surveys \cite{zou21}
we found seven more GRS that are brightest cluster members,
namely J0406--4429, J2052--4813, J2226--4316, J2240--4724, J2301--4359,
and J2325--4153, and four more lower-ranked cluster members.

In conclusion, we find 13 of 181~GRGs ($\sim7\%$) to be the brightest 
cluster members and 9 ($\sim5\%$) are lower-ranked members of which J0320--4515 had
already been reported as a member of Abell~S0345. Similar fractions of GRS
in clusters were reported by \cite{dabhade20a, dabhade20b}.

\subsubsection{Remnant Radio Sources}   \label{sect:rrgs}

\begin{figure}[H] 
\includegraphics[width=6.75cm]{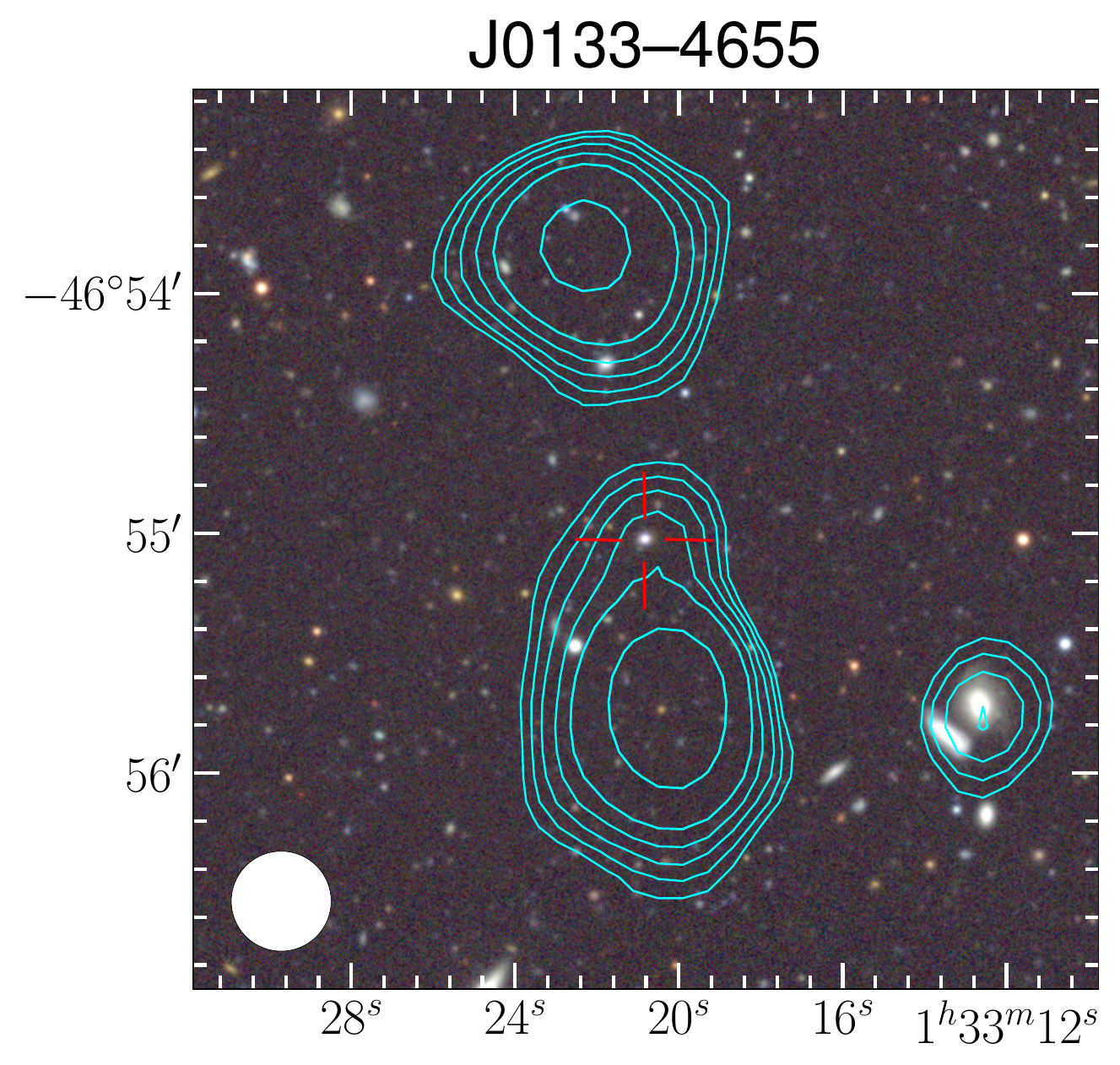}
\includegraphics[width=6.75cm]{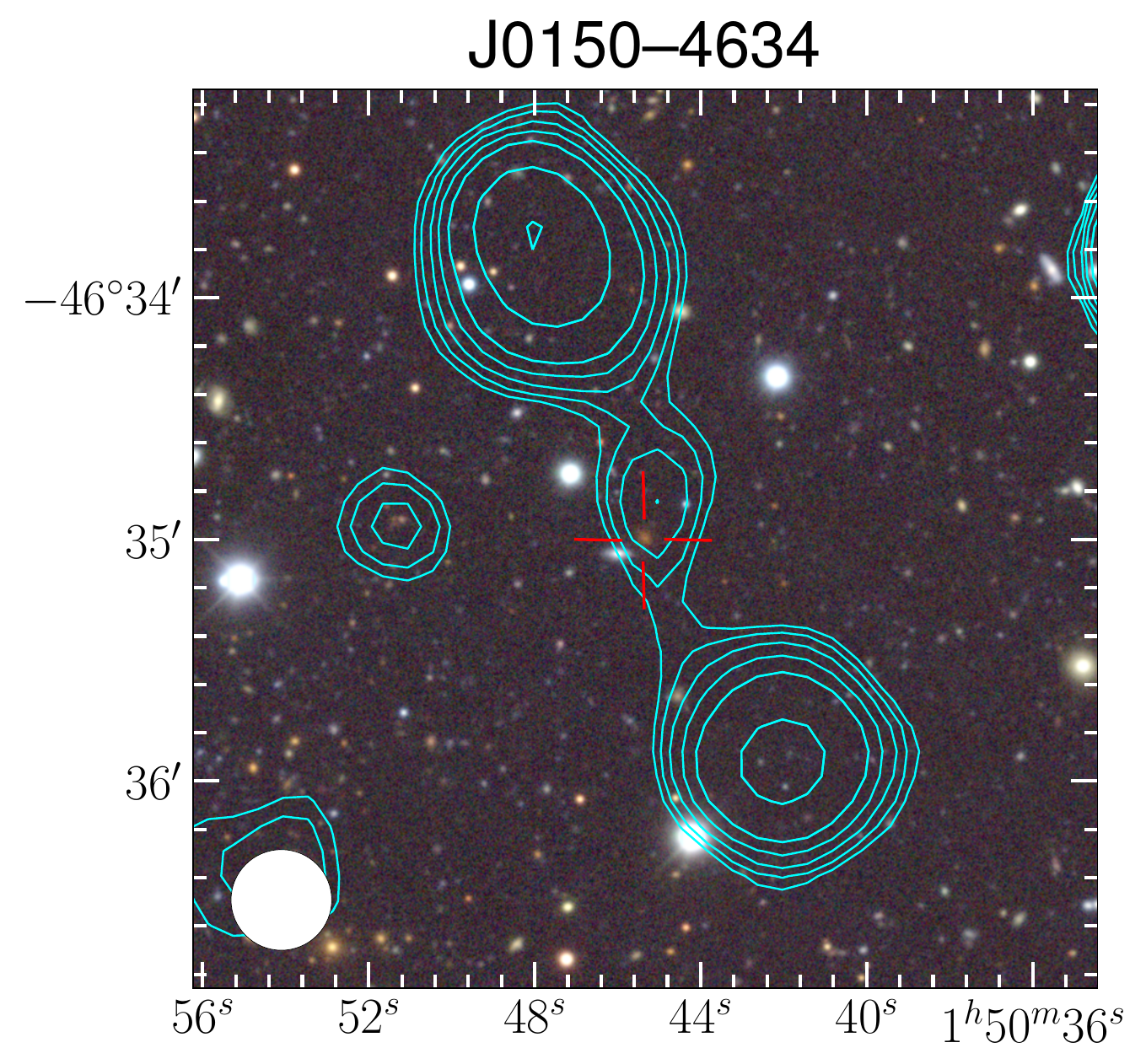}
\caption{Two examples of remnant-type GRS. On the left is J0133--4655 of
size 1.0\,Mpc, and on the right J0150--4634 of size 1.2\,Mpc with
indications of precession. Background image, contour levels,
and gapped red cross are as in Fig.~\ref{fig:2mpc}. }
\label{fig:rrgs}
\end{figure}

There is as yet no consensus about the criteria to classify
radio galaxies as of ``remnant" type \cite{mahatma18, jurlin20, quici21},
especially about whether to consider the presence of a radio core.  Here
we have made a crude attempt to assign the tag ``remnant" to objects
with very diffuse and little collimated lobes, and ``relic" to objects
with moderate indications of the latter, suggesting the absence, or at
least relative faintness, of any hotspots in these lobes. Note that
these tags were given as a proxy for the suspected age of the lobes,
\textit{independently} of the presence of a radio core, because (a)~the
presence of a radio core does not seem to have any relation with the
age of the lobes and (b) could be an indication of restarting activity
after a previous cycle of ``feeding" a pair of lobes with relativistic
particles. Thus the absence of a radio core is by no means an indicator
of the final ``death" of the radio activity. Fig.~\ref{fig:rrgs} shows 
two examples: at left is J0133--4655, hosted by the $r$=20.3\,mag galaxy 
DES~J013320.80--465501.4 at $z_{phot}$=0.65 with LLS=1.04\,Mpc. The galaxy 
near the centre of the South (S) lobe is
barely detected in WISE and is considered a chance superposition.
The fact that the S~lobe is closer to the host and $\sim$1.4 times more 
luminous than the North lobe, is consistent with statistical expectation
\cite{delarosa19}. At right is J0150--4634 (DES~J015045.33--463459.8) at
$z_{phot}$=0.88, of 1.15\,Mpc. The extended emission near the host is oriented
at an angle of $\sim30^{\circ}$ from the axis connecting the outer lobes,
suggesting the central engine is both precessing, as well as restarting its
radio activity.

\section{Discussion} \label{discussion}

The three previously reported GRS in our search area are
J0320--4515 of 1.8\,Mpc \cite{saripalli94}, and
J0116--4722 and J0213--4744 of 1.8 and 1.3\,Mpc, respectively,
published by \cite{subrahmanyan96}. Their redshifts range from
0.063 to 0.146 and their total 843-MHz flux densities from
SUMSS range from 2.1 to 6.7~Jy. Our new sample of GRS at high
Galactic latitude ($|b|\gtrsim25^{\circ}$) between Dec~$-40^{\circ}$ 
and $-50^{\circ}$ has hosts in the redshift range from 0.021 to $\sim$2.0
with total 888-MHz flux densities as low as 5.0\,mJy and a
median of $\sim$40\,mJy.

\subsection{Comparison of our Sample with previously published GRS} \label{sect:compar}

The literature compilation of GRGs by \cite{kuzmicz18} contains 223 objects
larger than 1\,Mpc, and after that four further large samples of GRGs and one
of GRQs were published. These contain (after correction of their LAS measures
and exclusion of a few spurious objects consisting of separate sources)
68, 48, 18 and 25 GRS from \cite{dabhade20a, dabhade20b, koziel20, brueggen21}
and 76 GRQs from \cite{kuzmicz21}. This makes a total of 458 GRS published,
neglecting here a few others that were reported in much smaller
numbers in various papers, sometimes not even mentioning their GRS nature.
Our new list of 178 GRS thus not only increases this number by $\sim$39\% from 
458 to 636, but is also the largest yet published individual list of newly found GRS.

Table~\ref{tab:compar} compares the six published samples (first
six rows) with our new sample (last row). Columns are (1) the surveys used and reference,
(2) their observing frequency, (3) the number of GRS $>$1\,Mpc published,
(4) the median LAS, (5) lowest and median total flux density of the GRS, (6) lowest and
median decimal logarithm of the radio spectral power at the frequency listed in col.~2,
(7) median redshift of the hosts, (8) the fraction of hosts with spectroscopic
redshift, and the fraction of hosts that are QSOs or candidates, and (9) the median LLS.
For comparison, for a spectral index of $\alpha=-0.8$, $log\,P_{150}$ is larger
by 0.62 and $log\,P_{1400}$ is lower by 0.15 than our $log\,P_{888}$.

\begin{table}[H]
\centering
\caption{\small Comparison of six previous lists of GRS with our new sample.}  \label{tab:compar}
{\footnotesize
\begin{tabular}{lcccccccc}
\toprule
    ~~~~(1)          &    (2)       &  (3)         &   (4)  &    (5)    &     (6)  &  (7)  &  (8)  &   (9) \\
Surveys, Reference   &  Freq.       & N of   & LAS         & S$_{min,med}$ & P$_{min,med}$ & z$_{med}$ & f$_{zsp}$,f$_{QSO}$ & LLS \\
                 &   GHz        & GRS      &   $'$    & mJy  &  W/Hz  &          &     &   Mpc   \\
\midrule
Literature \cite{kuzmicz18}  &  .8/1.4 & 223   & 6.7  &  9.2,154 & 23.4,25.5 & 0.23 & 0.89,0.17 & 1.41 \\
LoTSS      \cite{dabhade20a} &  0.15   &  68   & 3.5  &  7.0,130 & 24.7,26.2 & 0.54 & 0.55,0.12 & 1.28 \\
NVSS       \cite{dabhade20b} &  1.4    &  48   & 3.75 &   25,215 & 23.7,25.8 & 0.35 & 0.42,0.06 & 1.19 \\
NVSS,FIRST \cite{koziel20}   &  1.4    &  18   & 5.0  &  6.0,~40 & 23.5,24.8 & 0.22 &1.0~~,0.0~~& 1.18 \\
ASKAP      \cite{brueggen21} &  1.01   &  25   & 3.7  &  1.2,~22 & 23.3,25.4 & 0.6~~& 0.04,0.32 & 1.20 \\
NVSS,SDSS  \cite{kuzmicz21}  &  1.4    &  76   & 2.7  & ----,----& 25.3,26.0 & 0.82 & 0.97,1.00 & 1.22 \\
RACS, this\,work             &  0.888  & 178   & 3.1  &  5.0,~40 & 23.5,25.9 & 0.66 & 0.06,0.18 & 1.22 \\
\bottomrule
\end{tabular}}
\end{table}

Table~\ref{tab:compar} shows, not surprisingly, that the literature compilation
by \cite{kuzmicz18} has the largest median LAS and LLS, while ours is the one
with the second-smallest median LAS, only surpassed by \cite{kuzmicz21} who 
performed a dedicated search for extended
radio emission from spectroscopic QSOs from SDSS\,DR14Q. The latter is also 
the sample with the highest median host redshifts, followed by our present
sample. The sample with the lowest minimum and median flux density is from
ASKAP \cite{brueggen21} with its much higher sensitivity.

\begin{figure}[H] 
\includegraphics[width=6.75cm]{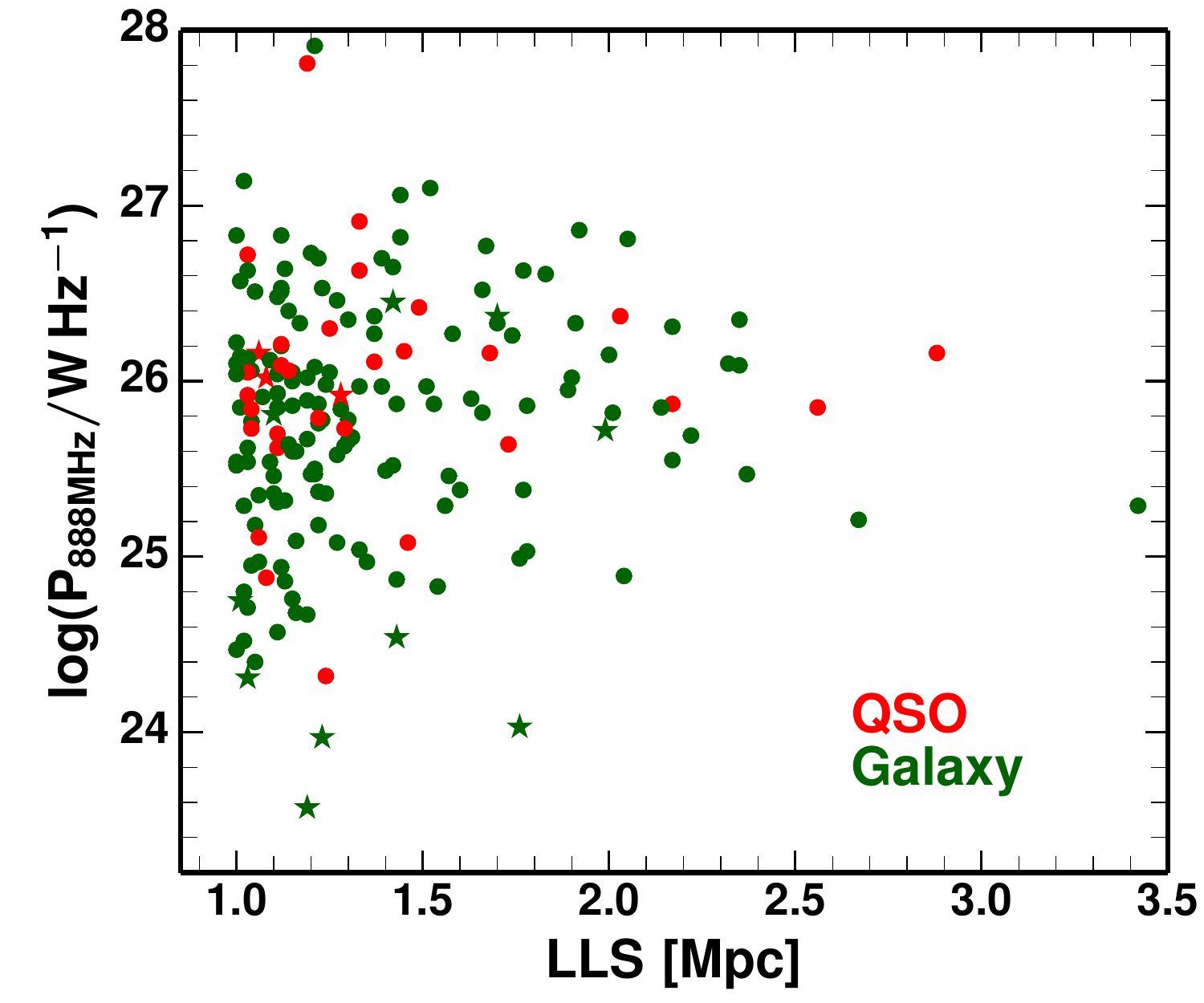}
\includegraphics[width=6.75cm]{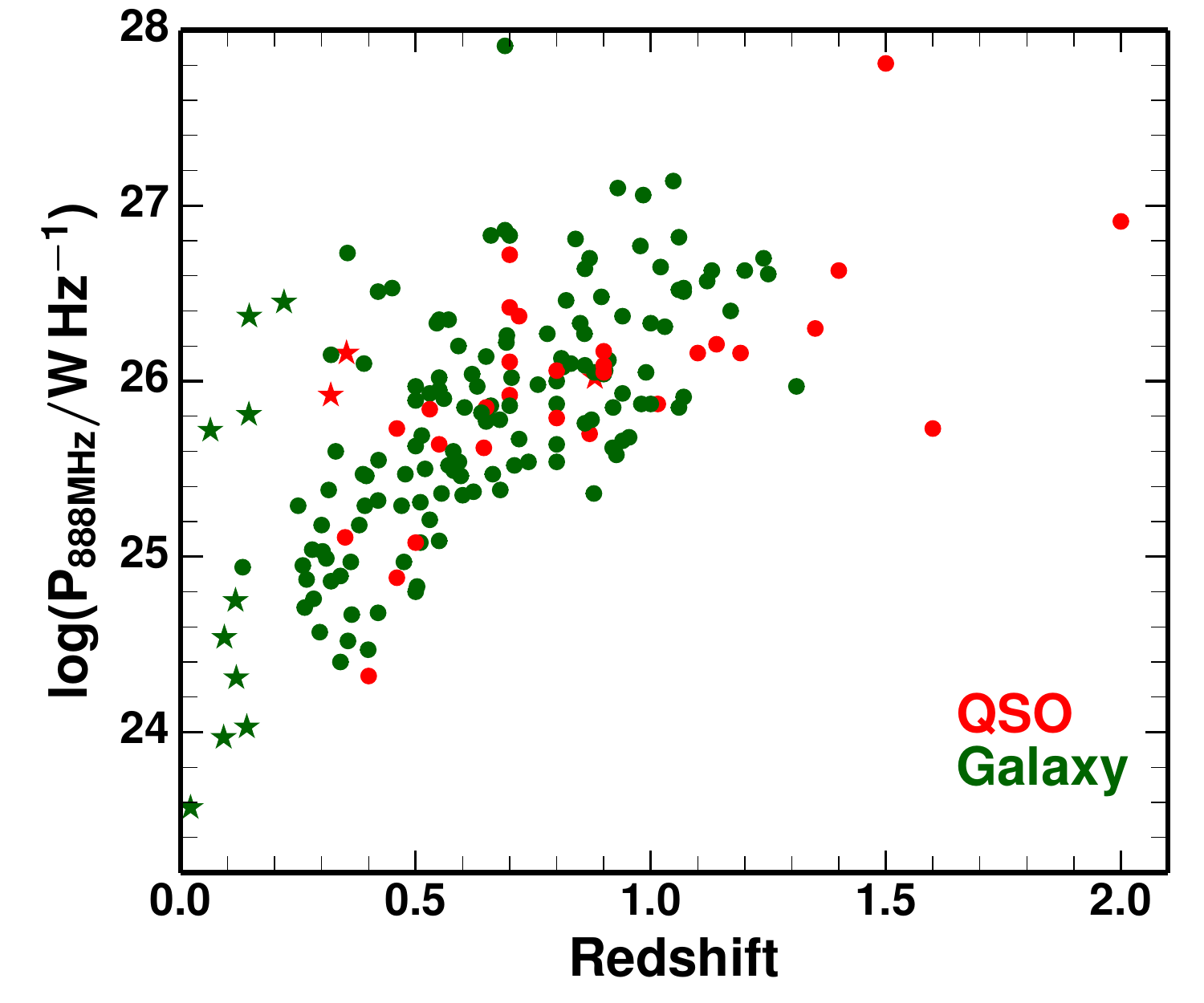}
\caption{Spectral radio power P$_{888}$ as function of LLS (left)
and as function of host redshift (right). Green filled circles are galaxies,
and red ones are QSOs or candidates with photometric redshifts, and
star symbols indicate objects with spectroscopic redshifts.}
\label{logP_lls_z}
\end{figure}

In Fig.~\ref{logP_lls_z} we present the distribution of spectral radio
power at 888\,MHz from our new GRS sample drawn from RACS, as  function
of LLS in the left panel, and as  function of host redshift in the right
panel. The left panel, also called the $P-D$ diagram is similar to
that in Fig.~5 of \cite{kuzmicz18} except for our sample being smaller,
such that any trend is washed out by the large dispersion in P$_{888}$.
The right panel shows a clear trend of P$_{888}$ rising with redshift
for galaxies, likely due to Malmquist bias. The panel is similar to that
of Fig.~4 of \cite{dabhade20a} except for the fact that our sample fills
in a lack of galaxies at $z\sim$1 which is due to the deeper DES\,DR1 images
used by us compared to the SDSS images used by \cite{dabhade20a}.
The dedicated search for GRQs among $\sim$526,000 SDSS~DR14 QSOs by
\cite{kuzmicz21} had shown a weak trend for radio power to increase with
redshift for GRQs with z$\gtrsim$0.5 out to z$\sim$2.5. Such a trend
cannot be seen in our much smaller sample due to the lack of
known spectroscopic QSOs in our search area.

\begin{figure}[H]
\includegraphics[width=13cm]{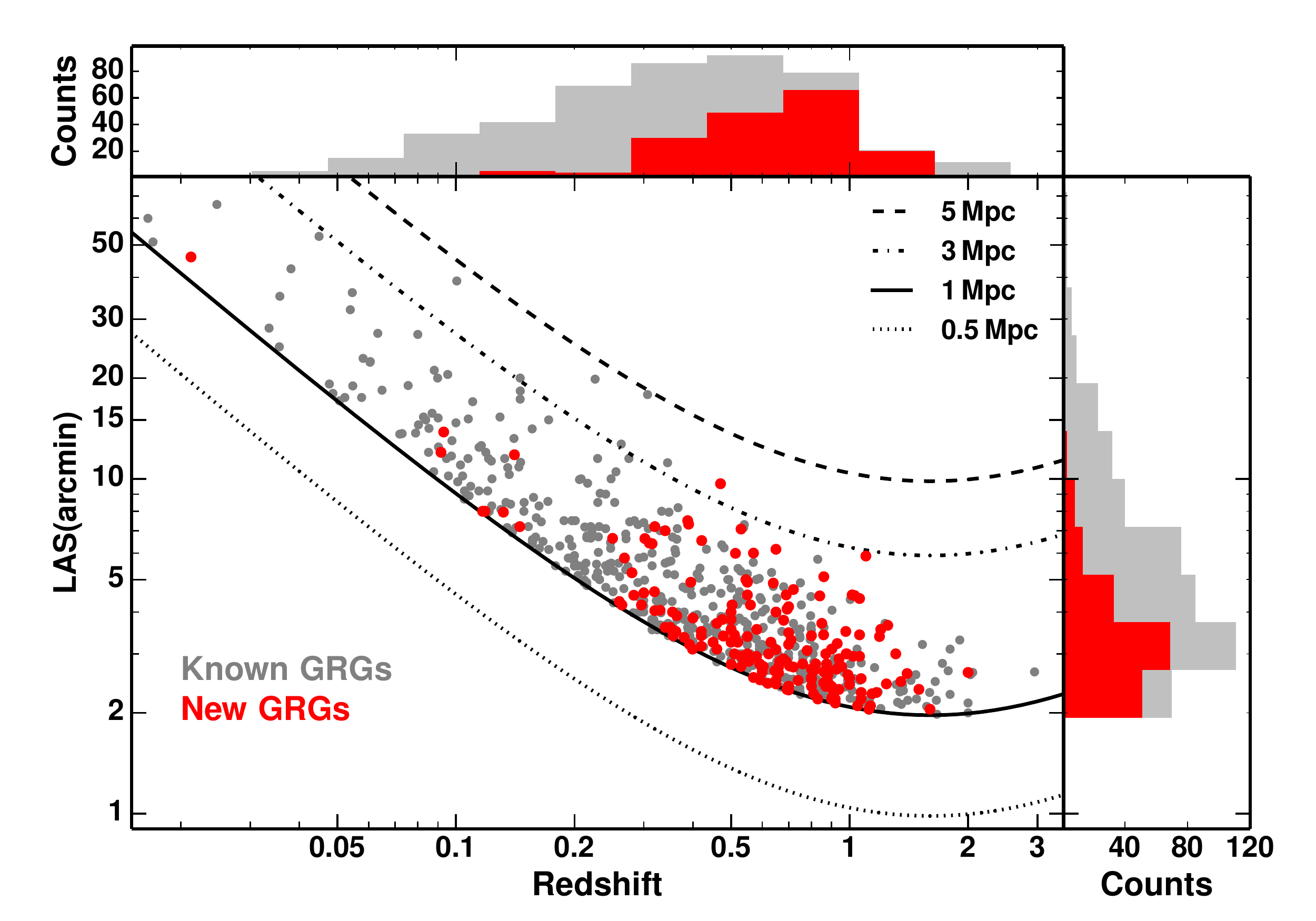}
\caption{The location of GRS in a LAS--redshift diagram. The grey points 
indicate the 458 GRS from the six major literature compilations, and the 
red points those in our new sample. The LAS of four ``standard rulers" 
of 0.5, 1, 3, and 5 Mpc are indicated by lines as described in the upper 
right corner of the diagram, assuming the cosmological 
parameters listed at the end of Sect.~\ref{intro}.}
\label{fig:las_z}
\end{figure}

\begin{figure}[H]  
\centering
\includegraphics[width=9cm]{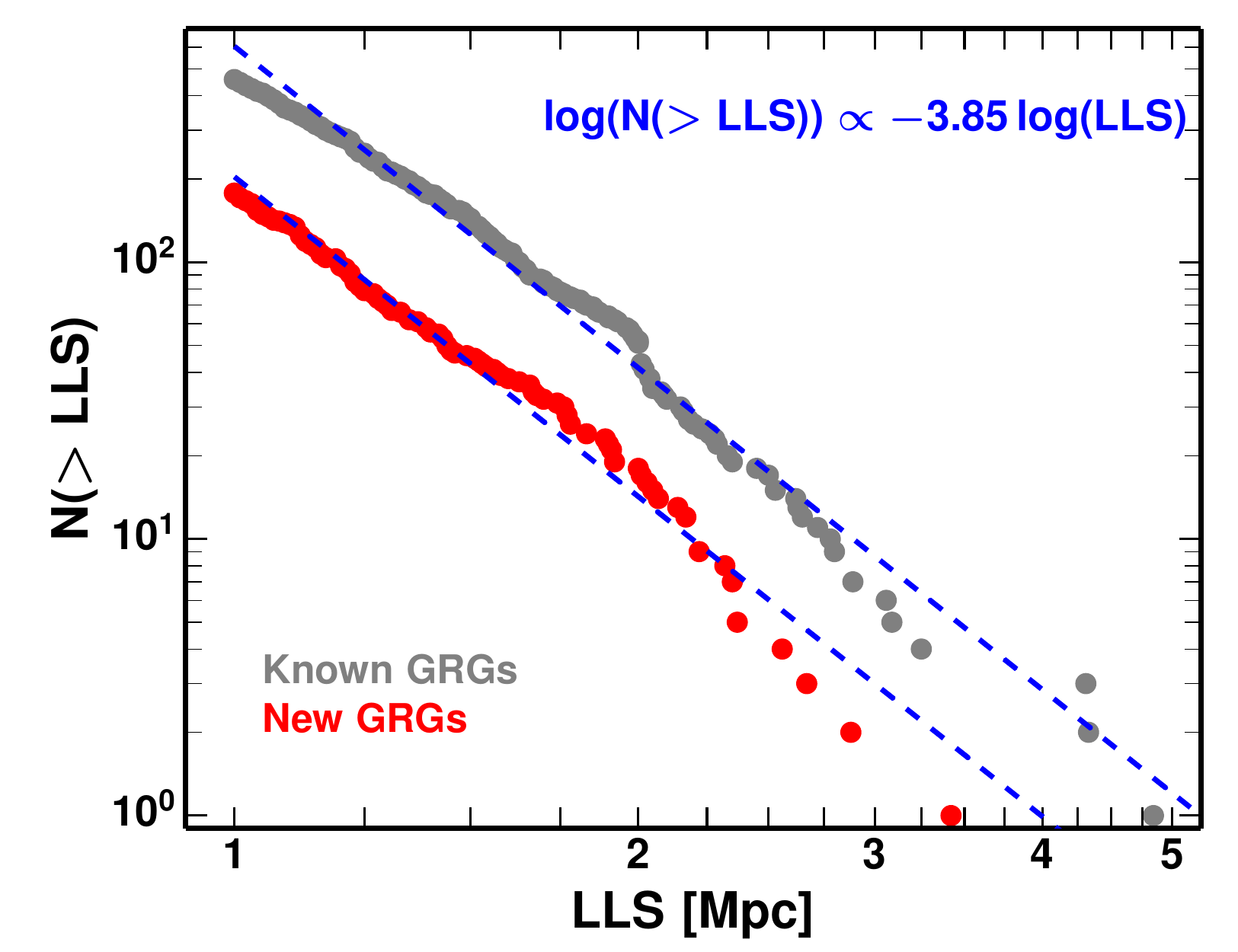}
\caption{The grey and red dots indicate the number of known GRS from the six publications
described in Sect.~\ref{sect:compar} and our 178 new GRS, respectively,
larger than a given LLS, as function of LLS. The slopes fitted to the log\,N(>LLS)--log\,LLS 
relation for the known and new GRS are $-3.86\pm0.04$ and $-3.84\pm 0.07$, respectively, 
and the dashed blue line is the average fit to both. The 
best-fit slope for the total population (not shown here) is $-4.02\pm0.04$.}
\label{fig:lls_cumul}
\end{figure}

In Fig.~\ref{fig:las_z} we show the location of our newly found 178 GRS in the
$LAS-z$ diagram, compared to the previously published 458 GRS, with
reference lines drawn for ``standard rulers" of four different sizes. The fact that
no GRS larger than 5\,Mpc has yet been found suggests this
may be a physical limit. The diagram also shows that GRS become much
more numerous for smaller LLS, but for LLS$<$1\,Mpc this has not been 
quantified in literature as yet.  Fig.~\ref{fig:lls_cumul}
shows the very rapid decrease of the number of GRS larger than a given
LLS as function of LLS. The data in Fig.~\ref{fig:lls_cumul} actually
show a flatter slope of this relation of about $-3.3$ below $\sim$2\,Mpc,
while above that size the slope is steeper than $-4$. Below 1\,Mpc this
slope becomes flatter: the number of $\sim$300 GRS found by us
with LLS=0.7--1.0\,Mpc, plus $\sim$400 of 0.5--0.7\,Mpc, suggest slopes 
of $-2.8$ and $-1.8$, respectively, for these latter size ranges.
For the literature compilation by \cite{kuzmicz18} with 
136 GRS of 0.7$<\rm{LLS}<1.0$\,Mpc the slope in this LLS range is much
flatter ($-1.4$), indicating that little attention had been paid to
these smaller sources in literature prior to this compilation.

\subsection{Multi-wavelength Counterparts of GRS hosts from RACS} \label{sect:multiwave}

The use of the deep multi-band optical images from the Dark Energy Survey
(DES\,DR1 \cite{abbott18}) brings a clear improvement over the use of previous
shallower optical surveys like the Digitized Sky Survey (DSS2, \cite{lasker96})
or SkyMapper (SMSS) DR1 \cite{wolf18}.
In fact, only 86 (43\%) of the 178 hosts of the new GRGs are listed
in the USNO\,B1.0 catalogue drawn from the DSS images \cite{monet03},
and only 35 (19.6\%) are found in the SMSS\,DR1 catalogue. In contrast,
175 of the 178 new GRG hosts are found in the DES\,DR1 catalogue \cite{abbott18}
and of the remaining three, J0600--4908 and J2331--4928
are listed in the  DESI\,DR9 catalogue (\cite{dey19},
\url{https://www.legacysurvey.org}), and J0131--4901 is
recognized on DESI\,DR9 images but not catalogued, so we estimated its redshift.

On the other hand, the WISE-based object catalogues are also very efficient 
in detecting GRG hosts: 163 (91.6\%) of the 178 new GRG hosts are detected in
AllWISE \cite{cutri13}, and 169 (96\%) are detected in CatWISE2020
\cite{marocco21}.

\subsection{Comparison of RACS with SUMSS} \label{sect:racs_sumss}

\begin{figure}[H] 
\includegraphics[width=6.75cm]{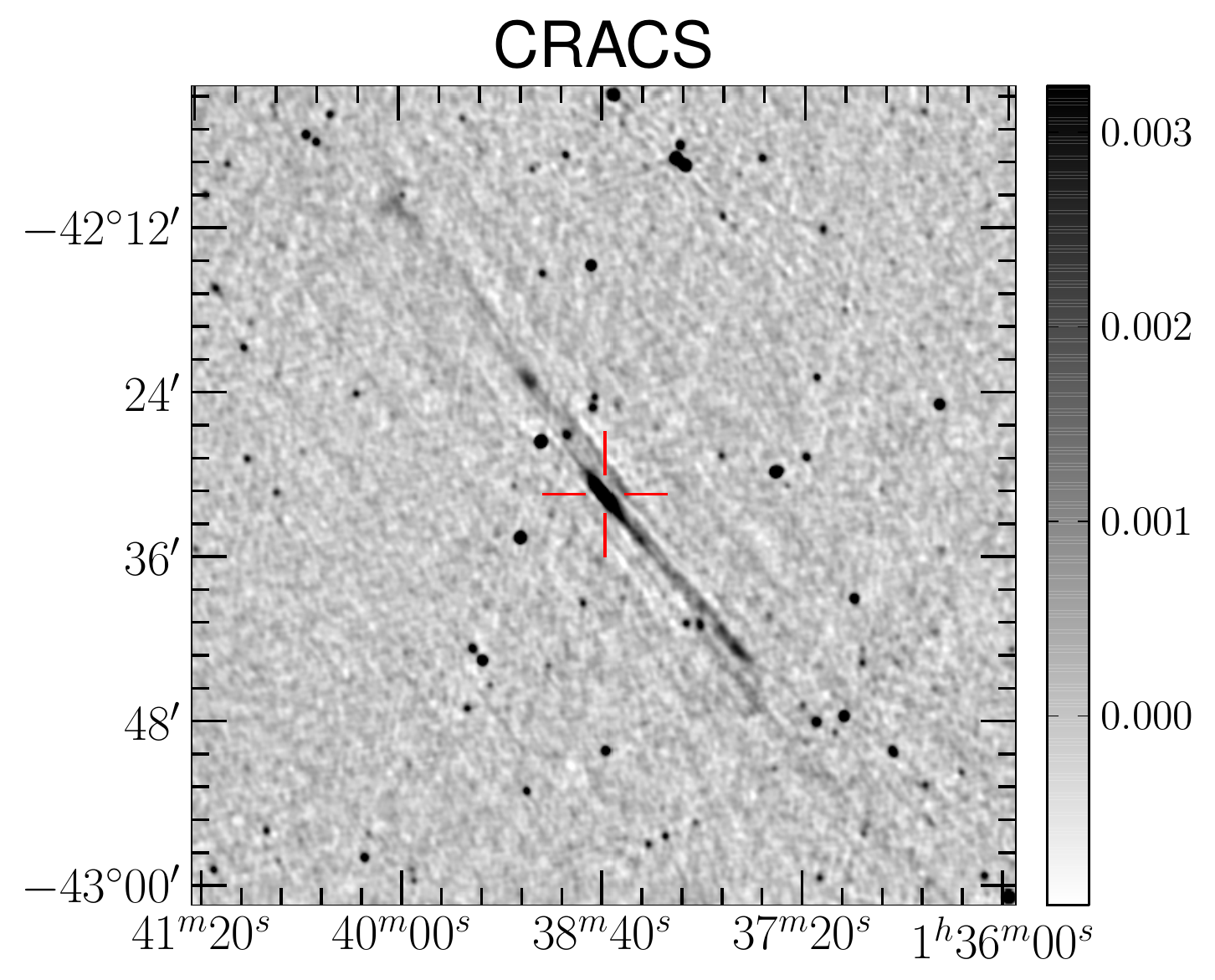}
\includegraphics[width=6.75cm]{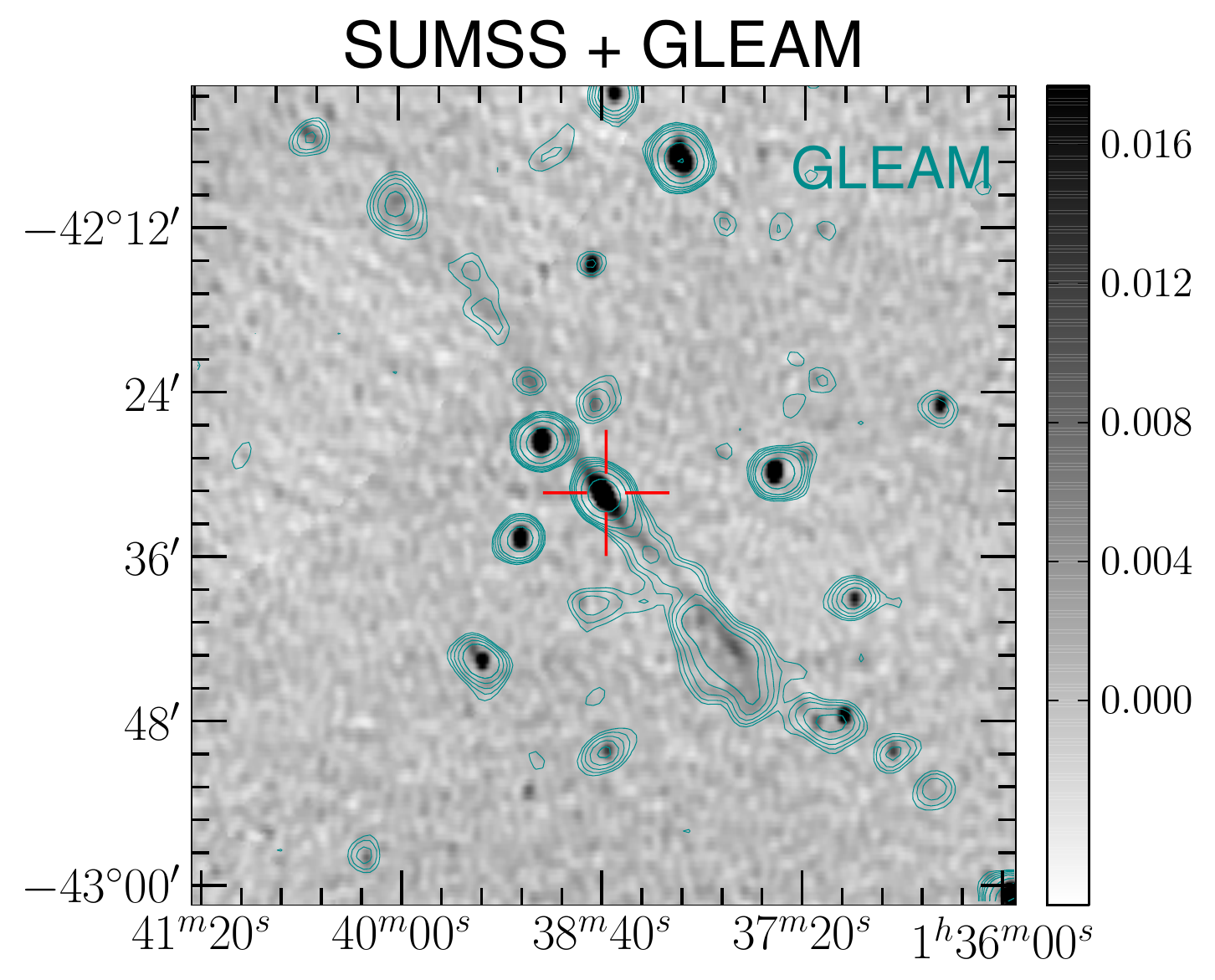}
\caption{Radio images of J0138--4231 (NGC\,641) from CRACS at 888\,MHz (left) and 
SUMSS at 843\,MHz (right), both in gray scale. The brightness levels in the 
scale bar on the right are in units of Jy\,beam$^{-1}$. The green contours
in the right panel are from the GLEAM 170--231\,MHz band at $\sim2'$ resolution,
plotted in the same sequence as in Fig.~\ref{fig:2mpc} but with rms=5.5\,mJy\,beam$^{-1}$.
The gapped red crosses at centre mark the location of the host.
See Sect.~\ref{sect:racs_sumss} for a discussion.}
\label{fig:ngc641racs_sums}
\end{figure}

The motivation for our effort to identify over 1400 extended
radio sources on RACS was its three times better angular resolution
of $\sim15''$ compared to that of SUMSS (45$''$).
However, we found that for sources with very large LAS of $\gtrsim15'$ and
very diffuse and low surface brightness lobes, SUMSS often
traces these sources with higher fidelity.
In Fig.~\ref{fig:ngc641racs_sums} we show one example of this by comparing
the CRACS and SUMSS images of J0138--4231, the nearby galaxy NGC\,641,
where the RACS image is affected by streaks parallel to the inner jets
and the SW lobe of the source, possibly due to the fact that ASKAP
has a shortest baseline of 22\,m \cite{hotan21} and RACS was observed
with a short exposure time of 15\,min, while SUMSS \cite{bock99} required
12-h integrations for the East-West array with virtually zero minimum spacings.
We added contours from GLEAM \cite{hurleywalker17} on top of the SUMSS image,
confirming the reality of the lobes.
Another example for which we found the SUMSS image of better quality
than RACS is J2226--4316, a 13.8$'$ wide GRS of LLS=1.4\,Mpc hosted
by the $z_{spec}$=0.0931 galaxy 2MASX~J22263358--4316356. The good
sensitivity of SUMSS for low surface brightness extended sources is
also borne out by the fact that 45 (25\%) of the 178 newly found GRS
were first seen on SUMSS images, only that the SUMSS angular resolution
did not allow their secure optical identification.

\section{Summary and Future Prospects}  \label{sect:summary}

We described the results of our search for radio sources extended
by more than $\sim1.4'$ on images of the RACS 888-MHz survey in a
contiguous high-Galactic latitude ($|b|>12.5^{\circ}$) area of 1059\,deg$^2$
between declinations of $-50^{\circ}$ and $-40^{\circ}$.  A subsequent
search for the optical host galaxies or quasars on color images of the
Dark Energy Survey DR1 resulted in the identification of $\sim$1000
such sources, and further use of available spectroscopic or photometric
redshifts revealed that 178 of these are giant radio sources (GRS)
larger than 1\,Mpc in projection and previously unreported in literature.
We thus increased the number of known GRS by $\sim$39\% from 458
to 636.  To our knowledge we present the largest sample of
newly discovered GRS exceeding 1\,Mpc in a single publication.
We also demonstrated that the depth and angular resolution of RACS,
together with deep optical/IR images, is perfectly suited for
the identification of large RGs.

About 18\% of the 178 new GRS are hosted by QSOs or QSO candidates which 
is similar to previously published GRS lists (except for the dedicated
search for GRQs by \cite{kuzmicz21}). At least 10\% of all GRS are hosted
by brightest or other cluster galaxies. This fraction is a lower limit,
since over half of our GRS lie at distances where cluster
catalogues are still very incomplete.

Finding 178 GRS in 1059\,deg$^2$ corresponds to a density of $\sim0.17$\,deg$^{-2}$
in this area and confirms that RACS images have the potential of revealing
$\sim$5000 GRS over its entire survey area (Dec$<+41^{\circ}$).
However, regions of low Galactic latitude as well as regions not covered
by the deepest optical surveys will yield a lower identification rate.
On the other hand, the fact that RACS is being repeated at two higher
frequencies ($\sim$1.2 and $\sim$1.6\,GHz) will increase its sensitivity
and potential for finding GRS.

We expect to double the present sample of GRS by inspecting the sky region
from RA=20$^h$ to 6$^h$20$^m$, and $-65^{\circ}<\rm{Dec}<-50^{\circ}$,
immediately south of that studied here, and also covered by DES,
and similar in size to the one just studied. Taken together, this promises
a sample of over 350\,GRS, plus $\sim$600 sources with LLS=0.7--1\,Mpc. 
However, the amount of work necessary to
find new GRS is clearly impractical and calls for machine learning (ML)
algorithms to find good candidates more efficiently. Samples
like the one presented in this work may serve as training sets for
such algorithms.

\vspace{6pt} 



\authorcontributions{Conceptualization, H.A.; visual inspection of radio 
and optical images,data curation, statistical analysis, H.A.; 
radio-optical overlays and all graphics, E.F.J.A; software development for
flux integration and determination of  equipartition 
parameters, A.G.W.; writing, H.A., E.F.J.A, A.G.W.
All authors have read and agreed to the published version of the manuscript.}

\funding{H.A. was partly funded by grant CIIC~174/2021 of
DAIP, Univ.\ de Guanajuato, Mexico.}

\dataavailability{All data used in this research are publicly available
from sites quoted in the present paper.  Table~\ref{tab:grglist}  will also be
made available at the CDS and through the VizieR service. The list of
radio galaxies smaller than 1\,Mpc identified in the present work may be
shared on reasonable request with researchers interested in collaborative 
projects.}

\acknowledgments{In this work we have used data from the ASKAP observatory. The Australian SKA Pathfinder is part of the Australia Telescope National Facility which is managed by CSIRO. Operation of ASKAP is funded by the Australian Government with support from the National Collaborative Research Infrastructure Strategy. ASKAP uses the resources of the Pawsey Supercomputing Centre. Establishment of ASKAP, the Murchison Radio-astronomy Observatory and the Pawsey Supercomputing Centre are initiatives of the Australian Government, with support from the Government of Western Australia and the Science and Industry Endowment Fund.

This research has made use of the VizieR catalogue access tool,
CDS, Strasbourg, France (DOI: 10.26093/cds/vizier). The original
description of the VizieR service was published in \cite{ochsenbein00}.

This research uses services or data provided by the
Astro Data Lab at NSF’s NOIRLab. NOIRLab is operated by the
Association of Universities for Research in Astronomy (AURA), Inc.\
under a cooperative agreement with the National Science Foundation.

The DESI Legacy Surveys consist of three individual and complementary projects:
the Dark Energy Camera Legacy Survey (DECaLS; Proposal ID \#2014B-0404;
PIs: David Schlegel and Arjun Dey), the Beijing-Arizona Sky Survey (BASS;
NOAO Prop. ID \#2015A-0801; PIs: Zhou Xu and Xiaohui Fan), and the Mayall
z-band Legacy Survey (MzLS; Prop. ID \#2016A-0453; PI: Arjun Dey). DECaLS,
BASS and MzLS together include data obtained, respectively, at the Blanco
telescope, Cerro Tololo Inter-American Observatory, NSF’s NOIRLab; the
Bok telescope, Steward Observatory, University of Arizona; and the Mayall
telescope, Kitt Peak National Observatory, NOIRLab. The Legacy Surveys
project is honored to be permitted to conduct astronomical research on
Iolkam Du’ag (Kitt Peak), a mountain with particular significance to
the Tohono O’odham Nation.

The Photometric Redshifts for the Legacy Surveys (PRLS) catalogue used
in this paper was produced thanks to funding from the U.S.\ Department
of Energy Office of Science, Office of High Energy Physics via grant
DE-SC0007914.

Use was made of data products from the Wide-field Infrared
Survey Explorer, which is a joint project of the University of California,
Los Angeles, and the Jet Propulsion Laboratory/California Institute of
Technology, funded by the National Aeronautics and Space Administration.

We also made use of data from the European Space Agency (ESA) mission
{\it Gaia} (\url{https://www.cosmos.esa.int/gaia}), processed by the {\it Gaia}
Data Processing and Analysis Consortium (DPAC,
\url{https://www.cosmos.esa.int/web/gaia/dpac/consortium}). Funding for the DPAC
has been provided by national institutions, in particular the institutions
participating in the {\it Gaia} Multilateral Agreement.

This research has made use of the NASA/IPAC Extragalactic Database,
which is funded by the National Aeronautics and Space Administration
and operated by the California Institute of Technology.

We are grateful to Karla Alamo Mart\'inez for help with extracting 
photometric redshifts
from DESI\,DR9, Ra\'ul F.\ Maldonado for his work in 2012 in spotting extended radio sources in SUMSS images, and to
Irina Andernach for her help with optical identifications of RACS sources.
We thank two anonymous referees for their useful comments.}

\conflictsofinterest{The authors declare no conflict of interest.}

\end{paracol}



\reftitle{References}

\end{document}